\documentclass[structabstract]{aa}
\usepackage{graphicx}
\usepackage{txfonts}
\usepackage{lscape}
\usepackage{longtable,lscape}
\usepackage{natbib}
\bibpunct{(}{)}{;}{a}{}{,} % to follow the A&A style
\def\Hb{${\rm H}_{\beta}$\,}

\def\apjl{ApJL}
\def\apjs{ApJS}

\def\aap{A\&A}
\begin{document}
\title{Early-type galaxies in the near-infrared: 1.5-2.4 $\mu$m
\\spectroscopy
\thanks{Based on observations made with ESO Telescopes at the La
Silla and Paranal Observatory under programmes 69.B-0311 and 077.B-0163.
}
\thanks{Table 5 is only available in electronic form at the CDS via
 anonymous ftp to cdsarc.u-strasbg.fr (130.79.128.5) or via
 http://cdsweb.u-strasbg.fr/cgi-bin/qcat?J/A+A/}
}

\author{M. Cesetti,
        \inst{1,2}
        V. D. Ivanov,\inst{1}
        L. Morelli,\inst{2}
        A. Pizzella,\inst{2}
        L. Buson,\inst{3}
        E. M. Corsini,\inst{2}\\
	E. Dalla Bont\`a,\inst{2}
        M. Stiavelli,\inst{4}
       \and
        F. Bertola\inst{2}
        }
\offprints{M. Cesetti}
\institute{European Southern Observatory, Ave. Alonso de
           C\`ordova 3107, Casilla 19, Santiago 19001, Chile\\
              \email{mcesetti@eso.org}
           \and
           Dipartimento di Astronomia, Universit\`a di Padova,
           vicolo dell'Osservatorio 3, I-35122 Padova, Italy\\
              \email{mary.cesetti@unipd.it}
           \and
           Istituto Nazionale di Astrofisica, Osservatorio 
           Astronomico di Padova, Vicolo dell'Osservatorio 5, 
           I-35122 Padova, Italy
           \and
           Space Telescope Science Institute,
           3700 San Martin Drive, Baltimore, MD 21218 USA\\
           }

   \date{Received July 03, 2008; accepted August 21, 2008}

  \abstract
{Near-infrared (hereafter NIR) data may provide complementary
information to the traditional optical population synthesis analysis
of unresolved stellar populations because the spectral energy
distribution of the galaxies in the 1-2.5\,$\mu$m range is dominated
by different types of stars than at optical
wavelengths. Furthermore, NIR data are subjected to less absorption
and hence could constrain the stellar populations in dust-obscured
galaxies.}
{We want to develop observational constraints on the stellar
populations of unresolved stellar systems in the NIR.}
{To achieve this goal we need a benchmark sample of NIR spectra of
``simple'' early-type galaxies, to be used for testing and calibrating
the outputs of population synthesis models. We obtained low-resolution
(R$\sim$1000) long-slit spectra between 1.5 and 2.4\,$\mu$m for 14
nearby early-type galaxies using SofI at the ESO 3.5-m New
Technology Telescope and higher resolution (R$\sim$3000)
long-slit spectra, centered at the Mg{\sc I} at $\sim$1.51\,$\mu$m for
a heterogeneous sample of 5 nearby galaxies
observed with ISAAC at Antu, one of the 8.2-m ESO Very Large
Telescope.}
{We defined spectral indices for CO, Na{\sc I},
Ca{\sc I} and Mg{\sc I} features and measured the strengths of
these features in the sample galaxies. We defined a new global NIR
metallicity index, suitable for abundance measurements in
low-resolution spectra.  Finally, we present an average NIR spectrum
of an early-type galaxy, built from a homogenized subset of our
sample.}
{The NIR spectra of the sample galaxies show great similarity and the
strength of some features does correlate with the iron abundance
[Fe/H] and optical metal features of the galaxies. The data suggest
that the NIR metal features, in combination with a hydrogen absorption
feature may be able to break the age-metallicity degeneracy just like
the Mg and Fe features in the optical wavelength range.}


   \titlerunning{Early-type galaxies in the near-infrared: 1.5-2.4
   $\mu$m spectroscopy} 
   \authorrunning{Cesetti, M. et al.}  
   \maketitle
%
%________________________________________________________________
\section{Introduction}

State-of-the-art space-based instrumentation can only resolve 
galaxies from the Local Group well and in the more distant objects we
usually can see only the tip of the red giant branch stars which is
rarely sufficient for population analysis. This leaves us with the
difficult task of trying to recover the stellar population of unresolved
galaxies from their integrated properties. The complex mix of stellar
populations found in most of them usually makes it possible to
constrain only the most recent generation of stars. However, such properties
as the stellar kinematics and the present-day metal content are
dominated by the overall star formation history, and together with the
well known age-metallicity degeneracy
\citep{fab73,oco86,wor94} they often lead to non-unique
solutions for the stellar populations.  In this respect, the Lick/IDS
system of indices pioneered by \citet{bur84} and \citet{fab85}, and
developed further by \citet{tra98, tra00, tra05}, has been
particularly successful for interpreting the integrated optical light
of galaxies.

However, new constraints are necessary to interpret more complicated
systems and one possibility is to widen the spectral range towards the
near-infrared (NIR) because the light in different wavebands is
dominated by different populations of stars. The NIR passbands are
dominated by light from older and redder star and therefore offer us
the possibility to study other stellar populations than is possible
with optical spectra alone. In addition, abundance determinations
through optical spectroscopy are not possible for heavily reddened
evolved stellar populations such as dusty spheroids or some bulge
globular clusters hidden by dust \citep[e.g.,][]{ste04}. NIR
spectroscopic observations could overcome these problem because the
extinction in the $K$-band is only one-tenth of that in the $V$-band.

Most of the previous work at NIR was focused on either active galactic
nuclei (AGNs) or objects with very strong star formation, including
recent surveys of ultra-luminous infrared galaxies
\citep{gol95,mur99,mur01,bur01}, luminous infrared galaxies 
\citep{gol97,reu07}, starbursts \citep{eng97,coz01}, Seyfert galaxies 
\citep{iva00,sos01,reu02,boi02}, LINERs \citep{lar98,alo00,sos01},
and interacting galaxies \citep{van97,van98}.

Relatively few NIR spectroscopic observations exist for ``normal''
galaxies. Only \citet{man01} provided low-resolution ($R\sim400$)
template spectra for galaxies of different Hubble types, including
some giant ellipticals. Such data, together with the corresponding
analysis is a necessary first step towards developing a system of
spectral diagnostics in the NIR because well-understood galaxies with
relatively simple star forming history will allow us to tune the NIR
population synthesis models. Recently, \citet[][hereafter S08]{sil08}
studied the stellar populations of eleven early-type galaxies in the
nearby Fornax cluster by means of $K$-band spectroscopy.

A few prominent NIR features were first studied by \citet{ori93} who
demonstrated that they represent a superb set of indicators for
constraining the average spectral type and luminosity class of cool
evolved stars. This conclusion was later confirmed by \citet{for00}
and \citet{iva04}. Furthermore, the same NIR features appear to be
promising abundance indicators \citep{fro01,ste04}.

Here we describe two new data sets of high quality NIR spectra of
ellipticals/spirals designed to provide a benchmark for future NIR
studies of unresolved galaxies: {\it (i)} low-resolution spectra
covering the range from 1.5 to 2.4\,$\mu$m that include strong
features such as CO, Na{\sc I} and Ca{\sc I} that are traditionally
studied, and {\it (ii)} moderate resolution spectra around the Mg{\sc
I} absorption feature at 1.51\,$\mu$m. This is the second strongest Mg
feature in the $H$- and $K$-band atmospheric windows (after the
feature at 1.71\,$\mu$m) and at zero redshift it is located in a
region of poor atmospheric transmission, making it difficult to
observe in stars. However, the redshift of external galaxies moves it
into a more transparent region \citep{iva01}, just the opposite of the
1.71\,$\mu$m Mg{\sc I} which becomes affected by the red edge of the
$H$-band atmospheric window.

We are only few years from the launch of the James Webb Space
Telescope \citep{gar06}, a space-based infrared telescope that will
have unprecedented capabilities. Therefore, in the near future the
application of NIR spectroscopy to the study of galaxy properties will
be limited not by lack of data but by our understanding of spectral
features at these wavelengths. Improving the characterization of the
NIR indices is a timely step in this direction.

The paper is organized as follows. The sample selection is discussed
in Sect.\,\ref{sec:Sample}. The NIR spectroscopic observations and
data reduction are described in Sect.\,\ref{sec:Obs}.  The definition
of the new NIR spectral indices and their measurements are given in
Sect.\,\ref{sec:Indices}. Results are discussed in
Sect.\,\ref{sec:Discus} and summarized in
Sect.\,\ref{sec:Results}.
%__________________________________________________________________
\section{The sample\label{sec:Sample}}
Our main sample consists of 14 nearby bright ($B_T$ $\leq$14.1\,mag)
and undisturbed spheroids. It was selected to cover a wide range in
luminosity and velocity dispersion and to have available Lick/IDS
line-strength indices, from the literature. Six of our galaxies are
giant ellipticals (NGC\,4472, NGC\,4621, NGC\,4649, NGC\,4697,
NGC\,6909, IC\,4296), five are intermediate-size ellipticals
(NGC\,4478, NGC\,4564, NGC\,4742, NGC\,5077, NGC\,5576), one is a
bright dwarf elliptical (NGC\,3641), and two are bulges of disk
galaxies (NGC\,4281, NGC\,4594; our spectra are dominated by their
spheroids).  The Mg$_2$ index spans the range between 0.21 and
0.34\,mag
\citep{ben93}, in agreement with other Mg$_2$ literature values,
and the central velocity dispersions are between about 130 and
370\,km\,s$^{-1}$. The galaxy properties are summarized in
Table\,\ref{tab:GalProperties1}.

Four galaxies in the sample exhibit evidence for weak nuclear
activity: NGC\,4594 and NGC\,5077 were identified as LINERs by
\citet{hec80}; NGC\,4472 is a weak Seyfert 2 \citep{ho97} 
and IC\,4296 shows radio and X-ray emission \citep{rin05}. Finally,
NGC\,4649 is a close companion of the giant spiral NGC\,4647 and it
may have undergone some tidal stripping
\citep{das99}.

The low resolution used for the observation of our sample is not
sufficient for characterizing the 1.51\,$\mu$m Mg{\sc I} feature, so
we selected from the ESO Science Archive higher resolution spectra
covering the region around the 1.51\,$\mu$m for a heterogeneous set of
galaxies. Although these observations were obtained for other
purposes, they provide us with a first glimpse into the behavior of
this feature. The properties of these galaxies are also described in
Table\,\ref{tab:GalProperties1}.

\begin{landscape}
\begin{table}
\begin{center}
\caption{Basic properties of the sample galaxies.}
\label{tab:GalProperties1}
\begin{tabular}{@{}r@{   }c c c l@{   }c c r r r r c c c c c c c@{}}
\hline
\hline
ID & Galaxy\hspace{3mm} & R.A.(2000)\hspace{1mm} & Decl.(2000)\hspace{1mm} & Type\hspace{1mm} & Redshift\hspace{1mm} & $\sigma$         & $B_T$ & $H_T$ & $K_T$ & \multicolumn{1}{c}{Age}   & Mg$_2$ & Mg$_2$ & \Hb     & Fe5335  & [$Z$/H] & [Fe/H]  & CaT     \\
   &                    &                        &                         &                  &                      & (km s$^{-1}$)    & (mag) & (mag) & (mag) & \multicolumn{1}{c}{(Gyr)} & (mag)  & (mag)  & (\AA) & (\AA) & (\AA) & (\AA) & (\AA) \\
(1)& (2)                & (3)                    & (4)                     & (5)              & (6)                  & (7)              & (8)   & (9)   & (10)  & \multicolumn{1}{c}{(11)}  & (12a)  & (12b)  & (13)    & (14)    & (15)    & (16)    & (17)    \\
\hline
\multicolumn{18}{c}{}\\
\multicolumn{18}{c}{Galaxies with low-resolution spectra}\\
1 \, &  IC\,4296	 & 13 36 38.85  &$-$33 57 59.3  & E         & 0.0124(a)  & 337.0(a)  & 11.61  & 14.56  & 14.30 &   5.2(j)  &  0.323  & 0.340(a)& 1.22(m)  & 3.03(m)  & 0.39(j) & \textellipsis & \textellipsis  \\
2 \, & NGC\,3641	 & 11 21 08.85  &  +03 11 40.2  & E\,pec    & 0.0058(b)  & 163.2(i)  & 14.10  & 15.92  & 15.17 &   6.6(k)  &  0.284  & 0.273(k)& 0.72(k)  & 2.39(k)  & \textellipsis & \textellipsis & 6.343  \\
3 \, & NGC\,3818	 & 11 41 57.50  &$-$06 09 20.0  & E5        & 0.0055(a)  & 187.5(a)  & 12.67  & 14.69  & 14.54 &   6.4(k)  &  0.315  & 0.322(a)& 1.67(k)  & 2.87(k)  & 0.37(o) &   0.16(o) & 6.242  \\
4 \, & NGC\,4281	 & 12 20 21.52  &  +05 23 12.4  &S0$^+$:\,sp& 0.0089(c)  & 230.5(h)  & 12.25  & 13.91  & 13.80 & \textellipsis &  0.314  & \textellipsis & \textellipsis & \textellipsis & \textellipsis & \textellipsis & \textellipsis \\
5 \, & NGC\,4472	 & 12 29 46.76  &  +07 59 59.9  & E2        & 0.0033(a)  & 309.7(a)  &  9.37  & 15.34  & 15.01 &  10.0(k)  &  0.306  & 0.331(a)& 1.37(k)  & 3.11(k)  & 0.26(o) &   0.06(o) & 6.119  \\
6 \, & NGC\,4478	 & 12 30 17.53  &  +12 19 40.3  & E2        & 0.0047(a)  & 153.0(a)  & 12.36  & 15.62  & 15.45 &   6.9(k)  &  0.253  & 0.260(a)& 1.67(k)  & 2.81(k)  & 0.30(o) &   0.16(o) & 6.810  \\
7 \, & NGC\,4564	 & 12 36 27.01  &  +11 26 18.8  & E         & 0.0038(a)  & 171.1(a)  & 12.05  & 14.74  & 14.66 &   8.3(k)  &  0.321  & 0.329(a)& 1.54(k)  & 2.81(k)  & \textellipsis &  \textellipsis & 6.325  \\
8 \, & NGC\,4594	 & 12 39 59.43  &$-$11 37 22.9  & Sa(s) sp  & 0.0034(a)  & 259.1(a)  &  8.98  & 12.22  & 12.22 &   9.5(k)  &  0.330  & 0.338(a)& 1.44(k)  & \textellipsis & \textellipsis &   0.39(o) & 6.361  \\
9 \, & NGC\,4621	 & 12 42 02.39  &  +11 38 45.1  & E5        & 0.0014(a)  & 232.0(a)  & 10.57  & 10.28  & 10.16 &   9.1(k)  &  0.328  & 0.345(a)& 1.39(k)  & 2.98(k)  & 0.07(o) & \textellipsis & 6.055  \\
10\, & NGC\,4649	 & 12 43 40.19  &  +11 33 08.9  & E2        & 0.0036(a)  & 368.5(a)  &  9.81  & 14.61  & 14.35 &  11.9(m)  &  0.338  & 0.347(a)& 1.29(k)  & 2.57(k)  & 0.29(o) &   0.06(o) & \textellipsis \\
11\, & NGC\,4697	 & 12 48 35.70  &$-$05 48 03.0  & E6        & 0.0041(a)  & 171.7(a)  & 10.14  & 13.35  & 13.31 &   6.4(k)  &  0.297  & 0.296(a)& 1.69(k)  & 1.94(k)  & 0.06(o) &$-$0.03(o) & 6.185  \\
12\, & NGC\,5077	 & 13 19 31.66  &$-$12 39 25.0  & E3-4      & 0.0090(d)  & 255.9(i)  & 12.38  & 16.21  & 14.93 &  15.0(j)  &  0.295  & 0.324(j)& 1.83(m)  & 2.64(m)  & 0.12(j) & \textellipsis & \textellipsis \\
13\, & NGC\,5576	 & 14 21 03.60  &  +03 16 14.4  & E3        & 0.0050(e)  & 183.1(i)  & 11.85  & 14.94  & 14.74 & \textellipsis &  0.253  & 0.243(f)& 1.52(n)  & 3.11(l)  & \textellipsis &   0.02(o) & \textellipsis \\
14\, & NGC\,6909	 & 20 27 38.60  &$-$47 01 34.0  & E+:       & 0.0090(f)  & 128.7(f)  & 12.61  & 14.25  & 14.32 & \textellipsis &  0.208  & 0.214(f)& 2.17(p)  & 2.32(p)  & \textellipsis & \textellipsis & \textellipsis \\
\multicolumn{18}{c}{}\\
\multicolumn{18}{c}{Galaxies with high-resolution spectra}\\
15\, & Mrk\,1055         & 02 48 18.55  &$-$08 57 37.4  & S?           & 0.0366(v)  & \textellipsis & \textellipsis & 14.69  & 14.30 & \textellipsis & \textellipsis &  \textellipsis & \textellipsis & \textellipsis & \textellipsis & \textellipsis & \textellipsis \\
16\, & NGC\,1144         & 02 55 12.32  &$-$00 11 01.7  & E            & 0.0288(q)  & \textellipsis & 13.78  & 15.18  & 15.06 & \textellipsis & \textellipsis & \textellipsis &  \textellipsis & \textellipsis &  \textellipsis & \textellipsis & \textellipsis \\
17\, & NGC\,1362         & 03 33 53.08  &$-$20 16 57.7  & S0$^o$:\,pec & 0.0041(t)  &  91(s) & \textellipsis & 16.11  & 15.68 & \textellipsis & \textellipsis & \textellipsis & 0.223(f) &  \textellipsis & \textellipsis & \textellipsis & \textellipsis \\
18\, & NGC\,4472         & \multicolumn{16}{c}{see above}\\	            		     	                
19\, & NGC\,7714         & 23 36 14.28  &  +02 09 17.8  & Sb(s):\,pec  & 0.0093(p)  &  62(u) & 13.00  & 15.85  & 16.08 & \textellipsis & \textellipsis & \textellipsis & \textellipsis & \textellipsis & \textellipsis & \textellipsis & 4.7(u)\\
\hline
\end{tabular}
\end{center}
Columns: (1) Identification number, used in the plots, (2) Galaxy
name, (3-4) Coordinates, (5) Morphological type from
\citet[][hereafter RC3]{dev91}, except for Mrk~1055 and NGC~1144 from 
the Lyon Extragalactic Database (LEDA), (6) Redshift, (7) Central
velocity dispersion, (8) Total observed blue magnitude from RC3, (9)-(10)
Total observed H and K magnitude from \citet[][2MASS All-Sky
Catalog of point sources]{cut03}, (11) Age, (12)-(17) Optical
indices available in literature, (12a) the Mg$_2$ values from
\citet{ben93} and (17) Calcium Triplet from
\citet{cen03}.\\ 
References: (a)-\citet{smi00}, (b)-\citet{fal99},
(c)-\citet{bin85}, (d)-\citet{dac98}, (e)-\citet{den05},
(f)-\citet{weg03}, (g)-\citet{kob99}, (h)-\citet{dal91},
(i)-\citet{fab99}, (j)-\citet{ann07}, (k)-\citet{san07},
(l)-\citet{how05}, (m)-\citet{ram05}, (n)-\citet{tra98},
(o)-\citet{tra00}, (p)-\citet{oga08}, (q)-\citet{kee96},
(r)-\citet{lau89}, (s)-\citet{ber02}, (t)-\citet{oli95},
(u)-\citet{gar06}, (v)-\citet{lu93}.
\end{table}
\end{landscape}

%__________________________________________________________________
\section{Observation and data reduction\label{sec:Obs}}

\subsection{SofI low-resolution spectra}

Long-slit ($0\farcs6\times290''$) NIR spectroscopy of 14 galaxies was
performed with the SofI (Son of ISAAC) infrared spectrometer
\citep{moo98b} at the Nasmyth A focus of the ESO 3.5-m New Technology
Telescope (NTT) at the European Southern Observatory (ESO) in la Silla
(Chile). The detector is 1024$\times$1024 pixels Hawaii HgCdTe array
with 18.5-$\mu$m pixel size. It is read out in four quadrants and the
average quantum efficiency is $\sim$\,65\,\%.  We used the low
resolution grism ($R\sim1000$) providing spectral dispersion of
10.22\,\AA\, pixel$^{-1}$ and a coverage between 1.53\,$\mu$m and
2.52\,$\mu$m.

The smaller galaxies were observed in nodding mode, placing them at
two different positions along the slit, separated by 150$''$. In the
case of bigger galaxies, the telescope pointing was alternated between
the galaxy and a nearby clear sky region to obtain a good sky
sampling.  We randomly dithered the telescope within a 10$''$ box
along the slit around each position. This dithering, also known as
jitter
\citep[see e.g. SofI handbook or][]{Devillard1999}, helps us to remove bad
pixels, to improve the pixel sampling, and the flat-field correction
because on each image the object is placed on slightly different
position on the array.

Immediately before or after each galaxy spectrum we observed a B star
to measure the atmospheric transmission and the instrument response,
always in nodding mode and with a similar airmass to that of the
corresponding galaxy.  Finally, to calibrate the spatial distortions
we also obtained a sequence of stellar spectra placed on 19
equidistant positions along the slit.

The details about the spectroscopic observations are given in
Table\,\ref{tab:log1}.

\begin{table}
\caption{Log of the spectroscopic observations}
\label{tab:log1}
\begin{center}
\begin{tiny}
\begin{tabular}{@{}c@{ }c@{ }c@{}c@{}r@{}c@{}c@{}c@{}c@{ }c@{}}
\hline
\hline
ID\hspace{1mm}   & Galaxy  & \hspace{2mm}Obs. Date\hspace{2mm} & Exp.     & \hspace{3mm}Slit\hspace{1mm} & Galaxy   & \multicolumn{2}{c}{Standard Star} & Aper-  & S/N \\
     &         & YYYY-        & Time     & P.A.    & sec\,$z$ & Sp.Type          &  sec\,$z$      & ture   &  \\
     &         & MM-DD        & (sec)    & (deg)   &          &                  &                & (arcsec)  &  \\
(1)  & (2)     &    (3)       & (4)      & \multicolumn{1}{c}{(5)} & (6)      & (7)              & (8)            & (9)   & (10)  \\
\hline
\multicolumn{10}{c}{} \\
\multicolumn{10}{c}{Galaxies with low-resolution spectra} \\
1 \, &  IC\,4296 & 2006-04-17 & 1200 &      0.0 & 1.19 & B2V   & 1.19 &  6.8 &  85.5 \\
2 \, & NGC\,3641 & 2006-04-17 & 3600 &     70.0 & 1.20 & B2V   & 1.21 &  4.0 &  60.2 \\
3 \, & NGC\,3818 & 2006-04-17 & 3600 &    100.0 & 1.34 & B2IV  & 1.34 &  5.7 &  81.2 \\
4 \, & NGC\,4281 & 2006-04-17 & 2400 &    178.0 & 1.26 & B2V   & 1.26 &  4.0 &  73.5 \\
5 \, & NGC\,4472 & 2006-04-16 & 2000 &    155.0 & 1.32 & B5    & 1.32 & 17.0 &  83.7 \\
6 \, & NGC\,4478 & 2006-04-16 & 2400 &    140.0 & 1.39 & B5V   & 1.39 &  8.5 &  64.5 \\
7 \, & NGC\,4564 & 2006-04-16 & 1800 &    137.0 & 1.57 & B5    & 1.57 &  5.7 &  89.2 \\
8 \, & NGC\,4594 & 2006-04-16 & 1200 &      0.0 & 1.42 & B5V   & 1.42 &  5.7 &  54.7 \\
9 \, & NGC\,4621 & 2006-04-17 & 1200 &    165.0 & 1.33 & B2IV  & 1.33 &  5.7 &  85.2 \\
10\, & NGC\,4649 & 2006-04-16 & 2000 &    105.0 & 2.27 & B5III & 2.27 & 22.7 &  88.7 \\
11\, & NGC\,4697 & 2006-04-16 & 1400 &     70.0 & 1.33 & B5    & 1.33 &  8.5 &  85.7 \\
12\, & NGC\,5077 & 2006-04-18 & 3200 &     10.0 & 1.24 & B8IV  & 1.24 &  6.8 &  75.2 \\
13\, & NGC\,5576 & 2006-04-16 & 1500 &     95.0 & 1.65 & B2.5V & 1.65 &  5.7 & 102.7 \\
14\, & NGC\,6909 & 2006-04-17 & 2000 &     70.0 & 1.09 & B2V   & 1.09 &  2.8 &  52.5 \\
\multicolumn{10}{c}{} \\
\multicolumn{10}{c}{Galaxies with high-resolution spectra} \\
15\, & Mrk 1055 & 2002-08-07 & 1800 &    179.5 & 1.05 & G3V   & 1.04 &  2.9 &  78.0\\
16\, & NGC 1144 & 2002-08-09 & 1800 &    117.7 & 1.21 & G3V   & 1.68 &  1.5 &  52.0\\
17\, & NGC 1362 & 2002-08-19 & 1800 &    179.5 & 1.02 & B3V   & 1.23 &  1.5 &  67.0\\
18\, & NGC 4472 & 2002-05-19 &  900 &    179.5 & 1.22 & B3III & 1.54 &  5.9 &  71.0\\
19\, & NGC 7714 & 2002-07-19 & 1800 &    179.5 & 1.14 & B3III & 1.03 &  2.3 &  96.0\\
\hline
\end{tabular}
\end{tiny}
\end{center}
(1) Identification number, used in the plots, (2) Galaxy name, (3)
Observing Date, (4) Total exposition time, (5) Slit position angle,
(6) Airmass, (7) Standard star spectral type, (8) Standard star
airmass, (9) Length of the apertures used to extract 1-dimensional
spectra shown in Figs.\ref{fig:spectraH},\ref{fig:spectraK} and 
\ref{fig:MgI_H_band}, (10) Average signal-to-noise per 1/pixel 
(see \S \ref{sec:Indices}).
\end{table}

The spectra were reduced with the Image Reduction and Analysis
Facility (IRAF)\footnote{Image Reduction and Analysis Facility is a
general purpose software system for the reduction and analysis of
astronomical data, IRAF is written and supported by the IRAF
programming group at the National Optical Astronomy Observatories
(NOAO).}. The sky emission was removed either by subtracting the
corresponding image in each nodding pair or by subtracting from each
object image the average of the preceding and the succeeding sky
images, respectively for the two modes of observation described
above. The data were flat-fielded with screen flat spectra. Geometric
distortion correction was applied along both the spatial and
dispersion axes by using an arc frame and the grid of stellar spectra
mentioned above, respectively.  Simultaneously, the frames were
wavelength calibrated. We applied the dispersion corrections, and
removed the hot pixels and cosmic ray hits with the algorithm
described in \citet{pyc04}.  Then, we aligned the individual
2-dimensional spectra along the spatial direction ({\it i.e.\/}, along
the slit), to combine them together and to extract a 1-dimensional
spectrum of the galaxy with apertures as listed in
Table\,\ref{tab:log1}.

\begin{figure}[t]
\includegraphics[width=10.0truecm]{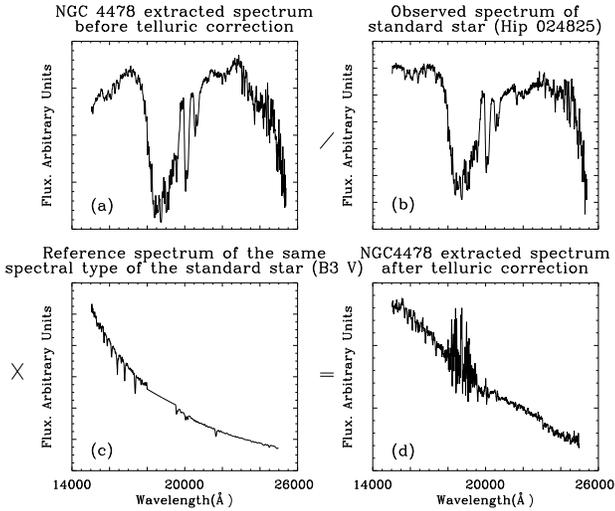}
\caption[]{The telluric absorption correction: the wavelength
calibrated 1-dimensional galaxy spectrum (a) is divided by the
wavelength calibrated 1-dimensional observed spectrum of the standard
star (b) and the result is multiplied by the reference of the same
spectral type of the standard star (c) from the library of
\citet{pic98}. The product (d) is the corrected galaxy spectrum. Note
the large scatter at $\sim$18000--19000\,$\mu$m region where the
Earth's atmosphere is practically opaque to the radiation.}
\label{fig:telluric}
\end{figure}

\begin{figure}%[h!]
\includegraphics[width=13.5truecm,angle=90]{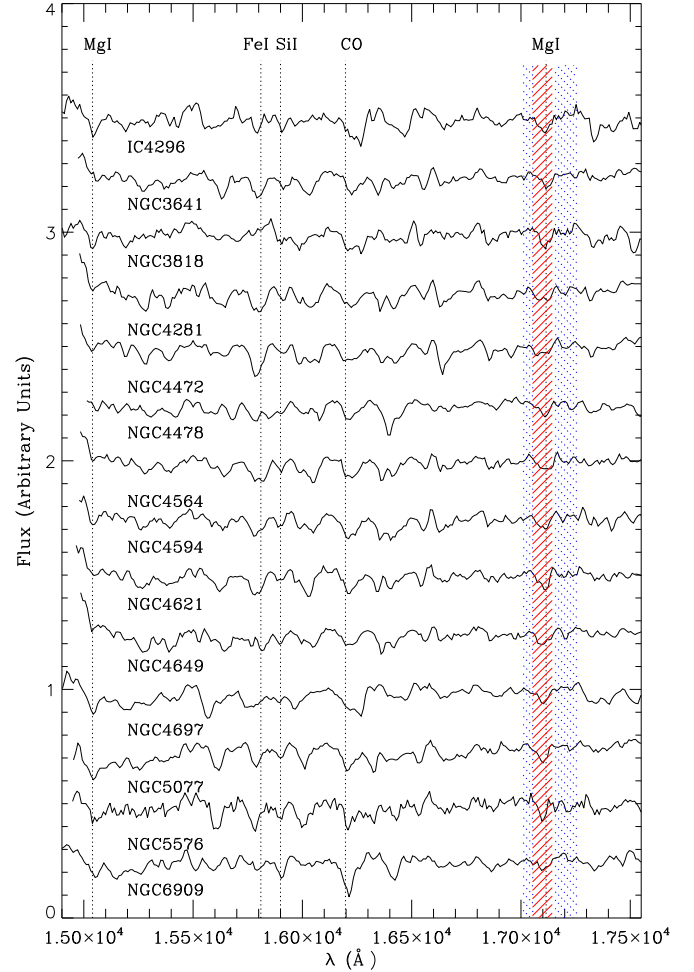}
\caption{$H$-band spectra of the sample galaxies, continuum-divided (i.e., 
all continua were normalized to 1) and shifted vertically for
displaying purposes by adding increasing shifts of 0.25. The position
of the main spectral features is shown with dotted lines. The
passbands of the Mg{\sc I} (1.71 $\mu$m) feature and its adjacent
continuum regions are marked with inclined solid and dotted lines,
respectively.}
\label{fig:spectraH}
\end{figure}

\begin{figure}%[h!]
\includegraphics[width=13.5truecm,angle=90]{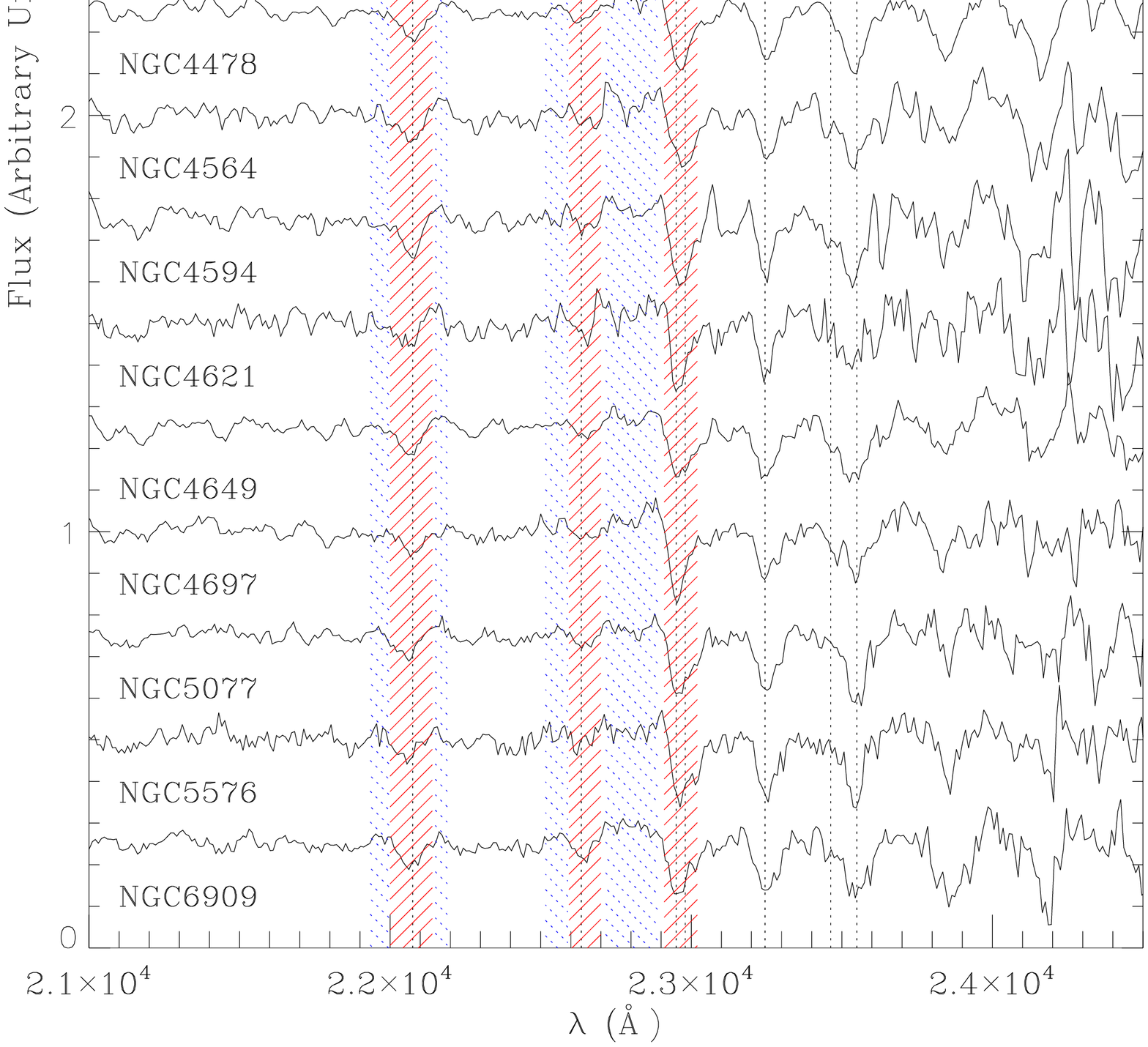}
\caption{$K$-band spectra of the sample galaxies, continuum-divided (i.e.,
 all continua were normalized to 1) and shifted vertically for
 displaying purposes by adding increasing shifts of 0.25. The position
 of the main spectral features is shown with dotted lines. The
 passbands of the Na{\sc I} (2.21 $\mu$m), Ca{\sc I} (2.26 $\mu$m),
 and CO (2.30 $\mu$m) features and their adjacent continuum regions
 are marked with inclined solid and dotted lines, respectively.}
\label{fig:spectraK}
\end{figure}

Finally, we removed the telluric absorption. To do this we first
created a sensitivity function by multiplying the observed spectrum of
the standard by the intrinsic spectrum of a star with the
corresponding (or the closest available) spectral type from
\citet{pic98}. Some residual emissions were observed, indicating that
the spectral types of the standards may not be accurate. We fitted
these artificial features with a Gaussian, with the IRAF task SPLOT
and subtracted them from the sensitivity function. This prompted us to
omit the bluest part of the spectra ($\lambda\leq$1.65\,$\mu$m) from
further analysis because the standards are rich of Hydrogen lines that
modify the appearance of the galaxy spectra. Next, we divided the
galaxy spectra by this sensitivity function to restore the overall
continuum shape of the galaxy spectrum. This process, broken into
steps, is illustrated in Fig.\,\ref{fig:telluric} and more detailed
descriptions can be found in \citet{han96} and \citet{mai96}. The
final reduced spectra were taken to zero redshift (the values used are
listed in Table
\ref{tab:GalProperties1}) and are shown in Figs.\,\ref{fig:spectraH}
and
\ref{fig:spectraK}.

\subsection{ISAAC high-resolution spectra}

Five spectra of galaxies were obtained with ISAAC -- a NIR imaging
spectrometer \citep{moo98a} at Nasmyth B focus of Antu, one of the
8.2-m ESO VLT unit telescopes of ESO Very Large Telescope at Paranal
(Chile). It is equipped with a 1024$\times$1024 Hawaii Rockwell HgCdTe
array with a spatial scale of 0.146\,$''$pixel$^{-1}$. The long-slit
($0\farcs3 \times 120''$) medium (R$\sim$3000) resolution spectroscopy
mode was used. The wavelength coverage spans the range from
$\sim$1.48\,$\mu$m to $\sim$1.62\,$\mu$m and the central wavelength
was chosen according to the object redshift. A summary of the
observations is given in Table\,\ref{tab:log1}.

Similar to the SofI observations, telescope pointings ware alternated
between the galaxy and a nearby empty sky region, or if the dimension
of the galaxy were small enough, then by nodding among two positions
along the slit separated by about 20$''$. No jittering was
used. Standards were observed in nodding mode soon before or after the
galaxy, at a similar airmass.  A sequence of stellar spectra placed at
equidistant positions along the slit was obtained to calibrate the
spacial distortions, as in the case of the SofI data.

The reduction steps were carried out using IRAF and they are the same
as for the main data set, albeit with some complication due to the
lower signal to noise and longer exposure times. First, the individual
images were cleaned from cosmic rays using SPEC\_COSM,
a cosmic ray rejection script for long slit spectra developed by
\citet{vand01}. Second, bad pixels were masked out before the 
combination because the small number of images limited the ability of
the rejection algorithms to remove the bad pixels during the
combination of the individual images. Finally, the highly variable sky
background led to residuals in the sky-subtracted images which were
removed with the IRAF task BACKGROUND before the geometric distortion
correction and the wavelength calibration.

%____________________________________________________________________
\section{Definition of the NIR spectral indices\label{sec:Indices}}

The NIR spectral indices are usually defined with a central bandpass
covering the spectral feature of interest, and two other bandpasses at
the red and blue sides which are used to trace a local continuum level
through a linear fit to the mean values in both bands, as in
\citet{ali95}, \citet{fro01} and \citet{iva04}. When this is not
possible, for example because the feature is at the end of the
atmospheric window, then either one continuum band or a fit to the overall
shape of the continuum is used, as in \citet{kle86}.

Most of the earlier definitions are aimed at stars, where the spectral
features are typically only a few km\,s$^{-1}$ wide, however in
galaxies the intrinsic velocity dispersion broadens them to a few
hundred km\,s$^{-1}$. This forced us to define new spectral indices
with broader passbands (Table\,\ref{tab:bandDef},
Fig.\,\ref{fig:spectraH} and \ref{fig:spectraK}). The continuum
passbands were placed on regions clear of absorption features, and not
(or only weakly) affected by the telluric absorption. Furthermore, we
tried to have as wide continuum passbands as possible, to increase the
signal-to-noise of the measurements. The passband of each spectral
feature was defined to encompass the maximum extent of the feature
itself and to avoid, if possible, contamination by other features. For
instance, the passband of our CO (2.30 $\mu$m) index is wide but still
only includes the $^{12}$CO feature and there is no contribution from
the $^{13}$CO band head.

\begin{table}[t]
\caption{NIR Spectral indices.}
The central wavelength $\lambda_c$ and bandwidth $\Delta\lambda$ are
given for the central feature (Line) and the continuum bandpass at
shorter (Cont. 1) and longer (Cont. 2) wavelengths.
\label{tab:bandDef}
\begin{center}
\begin{tabular}{l l l l l l l}
\hline
\hline
Feature                 & \multicolumn{2}{c}{Line}  & \multicolumn{2}{c}{Cont. 1}& \multicolumn{2}{c}{Cont. 2}  \\
                        & \multicolumn{2}{c}{-----------------}& \multicolumn{2}{c}{-----------------}& \multicolumn{2}{c}{-----------------}\\
                        & $\lambda_c$&$\Delta\lambda$& $\lambda_c$&$\Delta\lambda$ & $\lambda_c$&$\Delta\lambda$\\
                        & (\AA)    & (\AA)       & (\AA)    & (\AA)        & (\AA)    & (\AA)       \\
\hline
Mg{\sc I} (1.50 $\mu$m) & 15040      &  80           & 14960      &  80            & 15120      &  80 \\
Mg{\sc I} (1.71 $\mu$m) & 17098      &  90           & 17033      &  46            & 17206      & 100 \\
Na{\sc I} (2.21 $\mu$m) & 22070      & 140           & 21965      &  62            & 22170      &  40 \\
Ca{\sc I} (2.26 $\mu$m) & 22647      & 106           & 22553      &  74            & 22802      & 172 \\
CO  (2.30 $\mu$m)       & 22990      & 160           & 22802      & 172            & --         & --  \\
\hline
\end{tabular}
\end{center}
\end{table}

We measured the equivalent width (EW) of the lines with respect to a
local continuum obtained by fitting a straight line to the ``clear''
parts of the spectrum:
\begin{equation} \label{eq:EWformula}
{\rm EW} = (1 - F_{{\rm feature}} / F_{{\rm cont}}) * \Delta\lambda
\end{equation}
where $F_{{\rm feature}}$ is the flux inside the feature window,
$F_{\rm {cont}}$ is the value of the local continuum linearly
interpolated (or extrapolated, if necessary) at the center of the line
and $\Delta\lambda$ is the width of the feature bandpass. The fluxes
$F_{{\rm feature}}$ and $F_{\rm {cont}}$ are normalized by the
bandpass widths. The measurements were performed following the
above recipe with a self-written IDL script. The
error-bar are derived in the same script by means of Monte Carlo
simulations on the measured spectrum considering as noise the RMS of
the normalized flux in the continuum region. The average
signal-to-noise calculated on the continuum passbands regions is
written in Table \ref{tab:log1}. The corresponding values of each
equivalent width with their relative errors for all galaxies in the
sample are given in Table\,\ref{tab:EWObs}.

\begin{table}
\caption{Equivalent widths of the newly defined NIR spectral indices for
the sample galaxies.}
\label{tab:EWObs}
\begin{center}
\begin{tiny}
\begin{tabular}{l@{  }c@{  }c@{  }c@{  }c}
\hline
\hline
Galaxy  \hspace{2mm} & Mg{\sc I}  & Na{\sc I} & Ca{\sc I} & CO    \\
                     & EW $\pm\sigma$(EW)     & EW$\pm\sigma$(EW) & EW$\pm\sigma$(EW)  & EW$\pm\sigma$(EW)  \\
                     & (\AA)      & (\AA)     & (\AA)     & (\AA) \\
\hline
 IC 4296 &  2.44 $\pm$ 0.40  &  5.74 $\pm$ 0.37  &  4.01 $\pm$ 0.35  &  15.48 $\pm$ 0.63  \\
NGC 3641 &  2.54 $\pm$ 0.43  &  4.48 $\pm$ 0.40  &  1.88 $\pm$ 0.56  &  14.28 $\pm$ 0.57  \\
NGC 3818 &  3.23 $\pm$ 0.20  &  5.64 $\pm$ 0.19  &  2.67 $\pm$ 0.53  &  13.63 $\pm$ 0.65  \\
NGC 4281 &  3.07 $\pm$ 0.26  &  3.35 $\pm$ 0.28  &  3.08 $\pm$ 0.59  &  16.00 $\pm$ 0.80  \\
NGC 4472 &  3.04 $\pm$ 0.44  &  6.11 $\pm$ 0.18  &  2.72 $\pm$ 0.17  &  14.01 $\pm$ 0.63  \\
NGC 4478 &  3.50 $\pm$ 0.29  &  4.45 $\pm$ 0.43  &  1.91 $\pm$ 0.81  &  14.29 $\pm$ 1.00  \\
NGC 4564 &  2.97 $\pm$ 0.36  &  6.69 $\pm$ 0.15  &  2.25 $\pm$ 0.41  &  14.84 $\pm$ 0.71  \\
NGC 4594 &  3.24 $\pm$ 0.39  &  7.35 $\pm$ 0.35  &  3.01 $\pm$ 0.40  &  14.65 $\pm$ 1.12  \\
NGC 4621 &  2.73 $\pm$ 0.23  &  6.25 $\pm$ 0.11  &  2.79 $\pm$ 0.28  &  15.09 $\pm$ 0.64  \\
NGC 4649 &  2.77 $\pm$ 0.40  &  6.49 $\pm$ 0.16  &  3.10 $\pm$ 0.25  &  14.19 $\pm$ 0.60  \\
NGC 4697 &  3.14 $\pm$ 0.41  &  5.85 $\pm$ 0.20  &  3.10 $\pm$ 0.28  &  14.88 $\pm$ 0.58  \\
NGC 5077 &  3.72 $\pm$ 0.56  &  1.90 $\pm$ 0.15  &  2.78 $\pm$ 0.84  &  14.64 $\pm$ 1.06  \\
NGC 5576 &  2.37 $\pm$ 0.30  &  4.27 $\pm$ 0.15  &  1.75 $\pm$ 0.41  &  12.63 $\pm$ 0.70  \\
NGC 6909 &  3.60 $\pm$ 0.48  &  2.59 $\pm$ 0.46  &  2.72 $\pm$ 0.77  &  15.22 $\pm$ 0.94  \\
\hline
\end{tabular}
\end{tiny}
\end{center}
\end{table}

\begin{figure}%[h!]
\begin{center}
\includegraphics[width=4.4truecm]{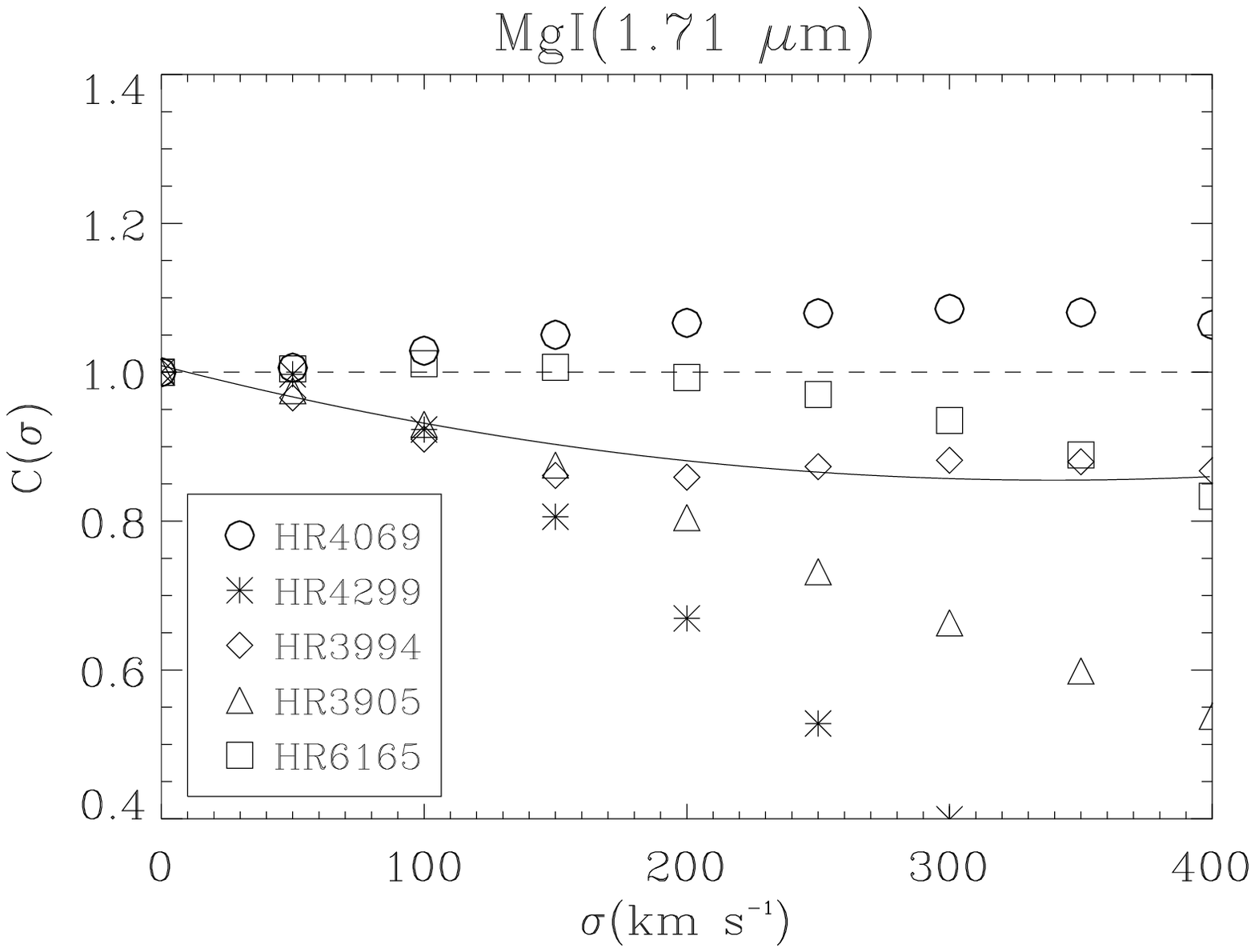}
\includegraphics[width=4.4truecm]{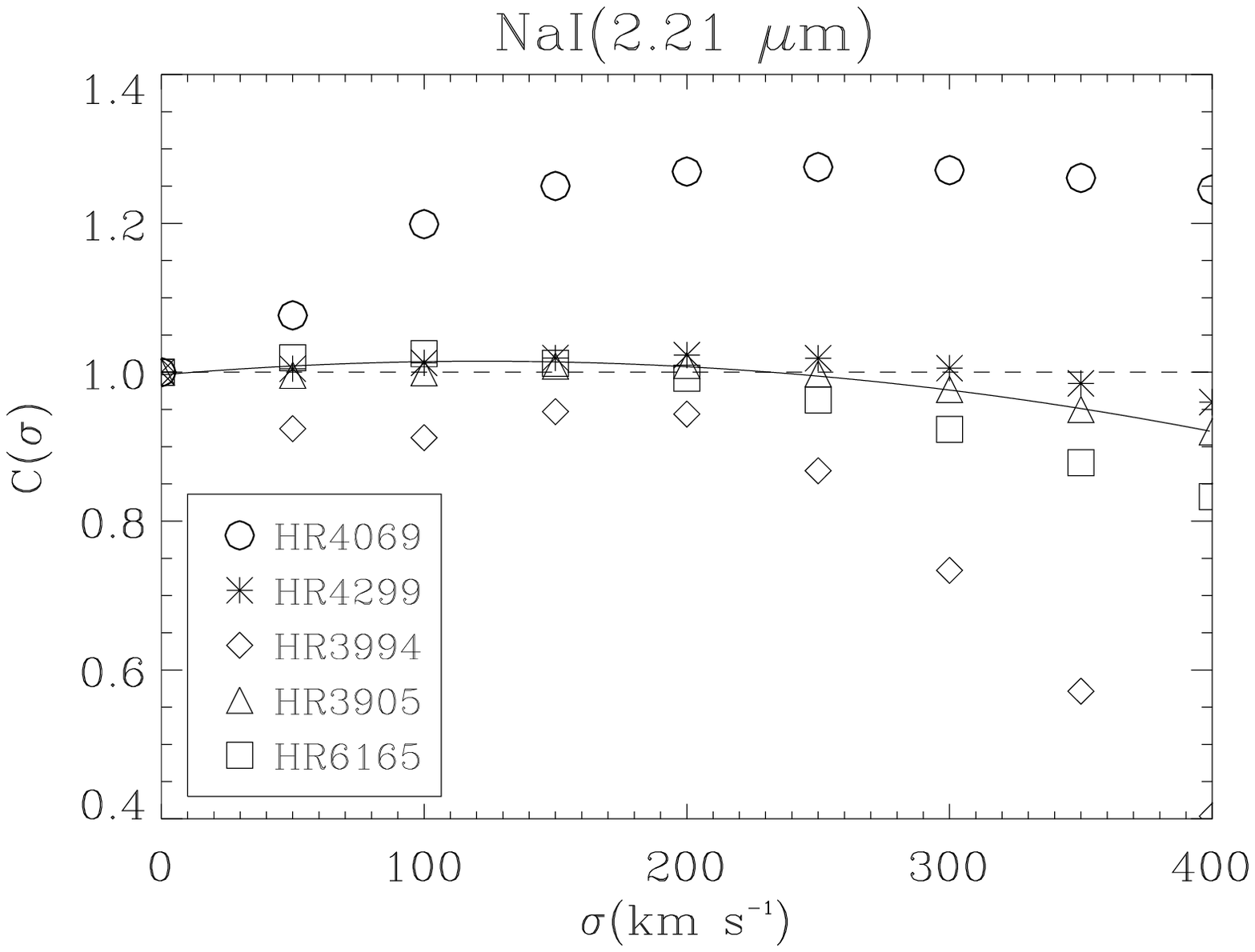}\\
\includegraphics[width=4.4truecm]{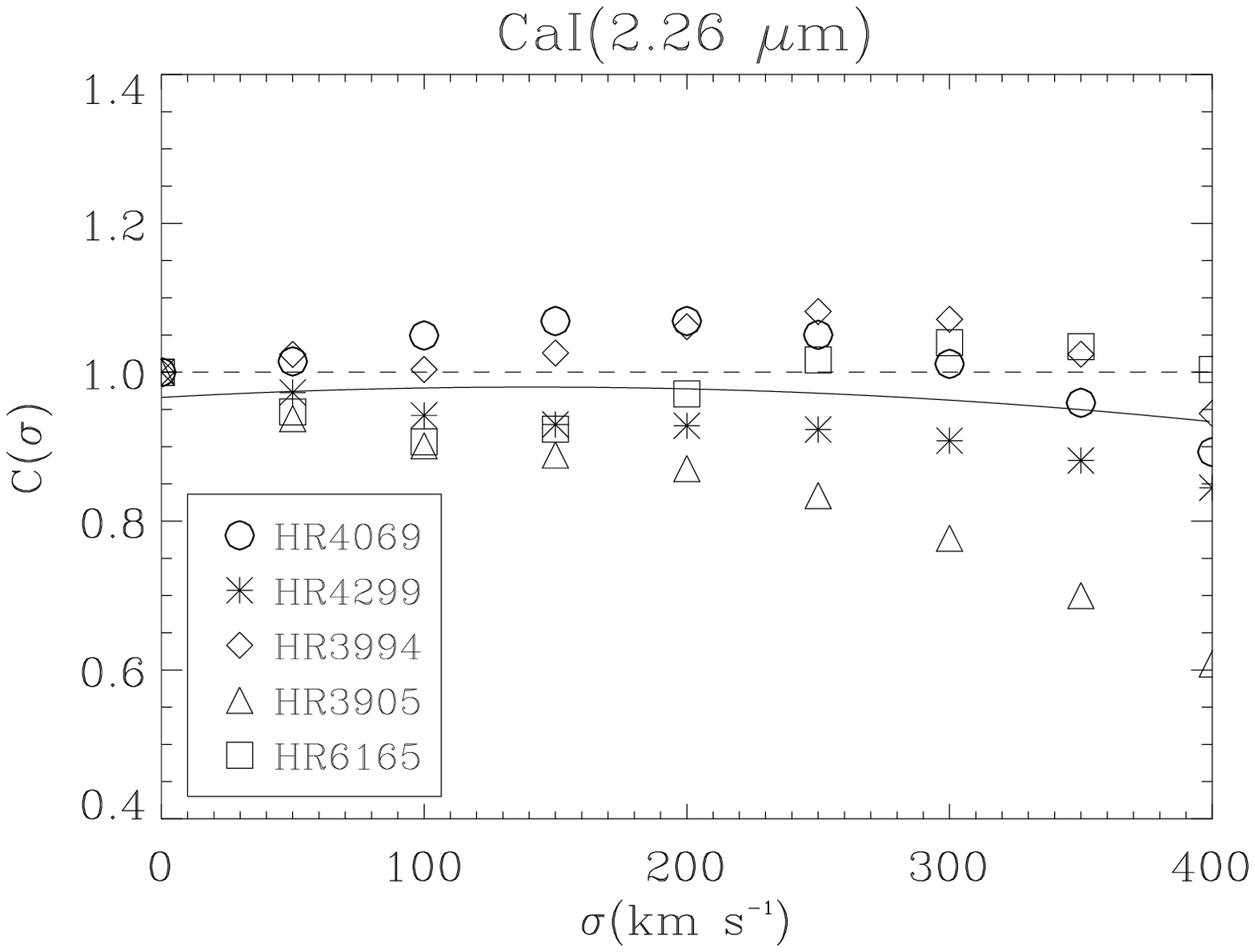}
\includegraphics[width=4.4truecm]{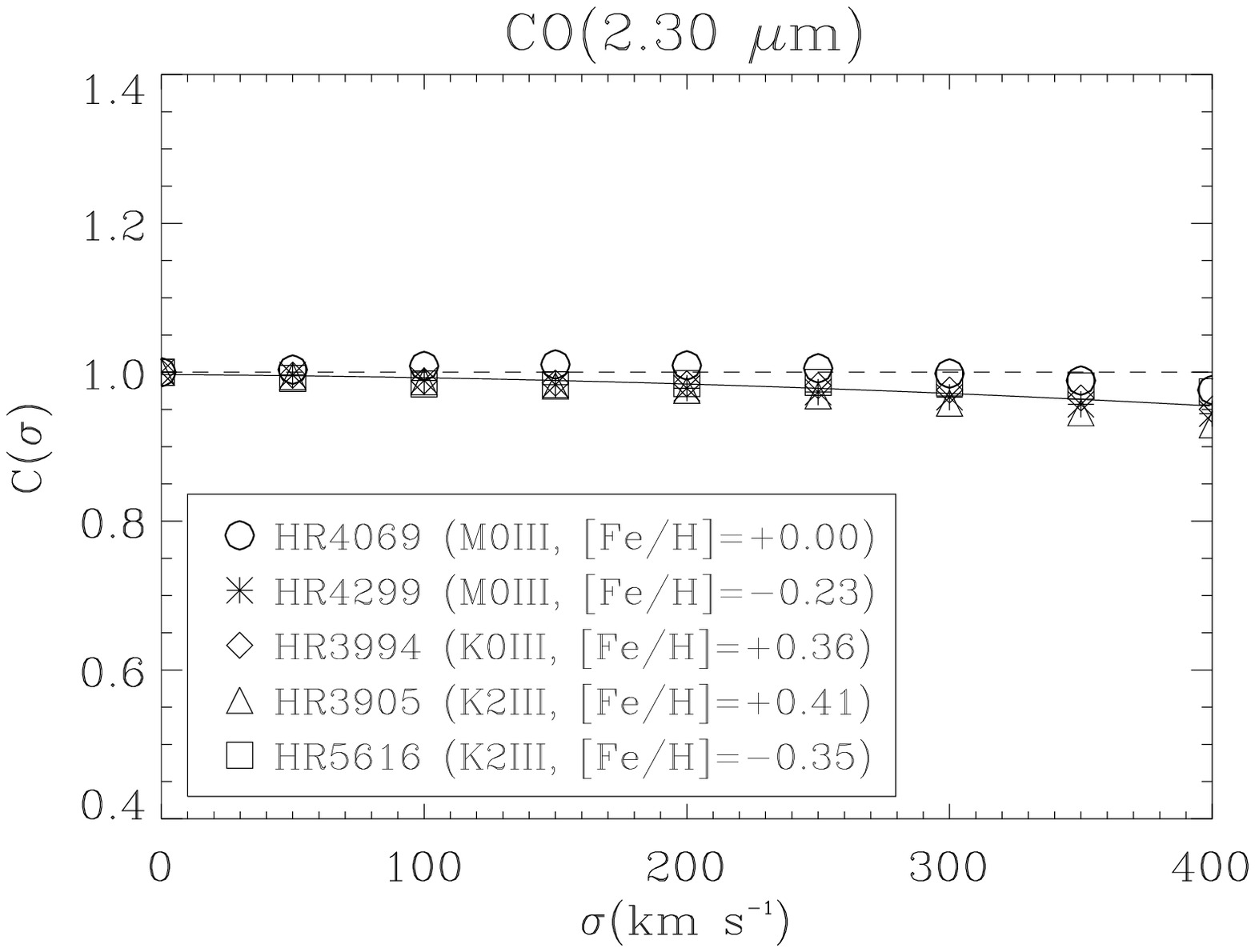}
\caption{Correction factors of the NIR spectral index for different 
values of velocity dispersion. They have been derived by comparing the
EW measured in the spectra of the labeled stars before and after
broadening them to a particular velocity dispersion.
The spectral types of the stars are in a range between KIII and MIII 
with a spread of [Fe/H] between -0.35 and +0.41. In each panel the solid line
is the 2nd order polynomial fitting the data.}
\label{fig:dispCorr}
\end{center}
\end{figure}

Our definitions have intentionally wide bandpasses to make them
intrinsically insensitive to the varying velocity broadening of
spectral features in the galaxy spectra. To verify this, we
measured the indices for broadened spectra of M and K type giants. We
choose stars in this spectral range because they are the most
representative of the early-type integrated galaxy spectra.  (These
spectra are a reasonable match to the observed galaxy spectra.) We
computed for them the correction factor $C(\sigma)$ for various
line-of-sight velocity distribution, as in S08:
\begin{equation}
C(\sigma)=\frac{{\rm EW}(\sigma=0)}{{\rm EW}(\sigma)}
\end{equation}
the EW of the feature is measured from the stars convolved with a
Gaussian with $\sigma$ given in brackets. We broadened the stellar
spectra to velocity dispersion up to 400\,km s$^{-1}$ in steps of
50\,km s$^{-1}$ and used templates for different spectral stars and
from different sources. The corrections for the different spectral
indices are shown in Fig.\,\ref{fig:dispCorr}. Some of them show a
significant scatter, depending on the adopted stellar
spectrum. However, the mean correction are typically small (i.e.,
$\la15\%$ in the worst case).

%____________________________________________________________________
\section{Discussion and Conclusions\label{sec:Discus}}

\subsection{General appearance of the spectra}

The spectra of our sample galaxies appear qualitatively similar 
in most of the NIR features except for a large spread in the Na{\sc I}
values is present.  This similarity is not surprising because we
selected mostly giant ellipticals and spheroids, which are all
relatively metal rich, have no significant recent star formation, and
only weak nuclear activity, if any. This conclusion agrees with
\citet{man01} who demonstrated that galaxies within the same Hubble
type have nearly identical NIR spectra.
On the other hand, the large spread of NaI, with respect to the 
observational errors, appears to be real, and suggests variety of star
formation and enrichment histories among the galaxies in our
sample. Two of the three galaxies with systematically weaker NaI
exhibit strong H$\beta$, indicating that the weak NaI might be related
to the presence of younger stellar populations (the third galaxy lacks
optical spectroscopy).  The strongest features are the CO absorption
bands in both the $H$ and $K$-band atmospheric windows. They originate
in K and M stars, as can be seen from the libraries of stellar spectra
described in
\citet{lan92}, \citet{ori93}, \citet{dal96}, \citet{mey98}, 
\citet{wal97} and \citet{for00}. A number of weaker metal absorption 
features are visible as well: Si{\sc I} at 1.589\,$\mu$m, Mg{\sc I} at
1.711\,$\mu$m, Na{\sc I} at 2.206 and 2.209\,$\mu$m and Ca{\sc I} at
2.261, 2.263, and 2.266\,$\mu$m. They are also present mostly in cool
stars \citep[i.e.][]{kle86,ori93,wal97,for00,mey98}. As
mentioned above, we only consider the redder features with
$\lambda\geq$1.65$\mu$m.

Not surprisingly, our galaxies show no Fe{\sc II} at 1.644\,$\mu$m
or H$_2$ lines line emission, usually associated with supernova
activity and supernova related shocks, respectively. There is also
no Br$\gamma$ emission. 

\subsection{Behavior of the NIR spectral indices}\label{sec:Behaviour}
Evolutionary synthesis models \citep[i.e.][]{wor94} are the usual
method to interpret the behavior of spectral features in galaxies, but
the simple comparison of these features with the galaxy parameters can
be informative too.  For example, the optical indices, Mg$_2$\,,
$<$Fe$>$ and \Hb show tight correlation with the central velocity
dispersion \citep[e.g.][]{ter81,ber98,meh03,mor08}, suggesting that
the chemical and the dynamical evolution of ellipticals are
intertwined.

\begin{figure}
\includegraphics[width=9truecm]{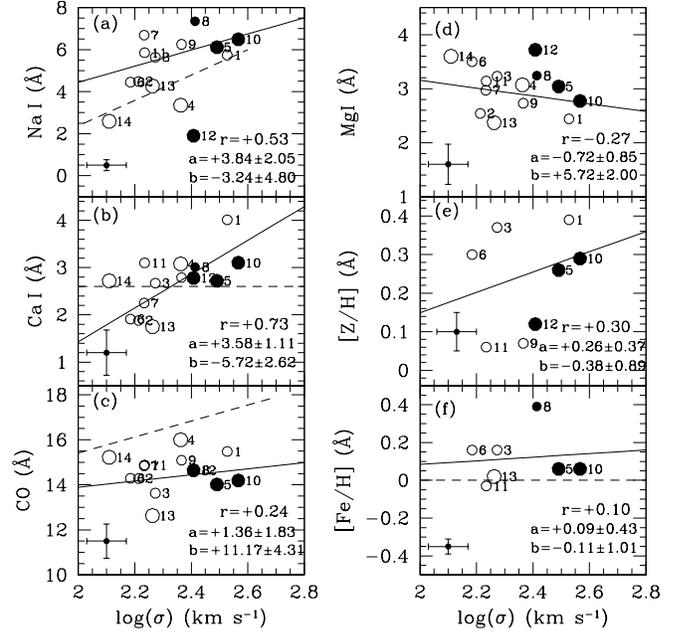}
\caption{The equivalent widths of the NIR spectral indices, iron abundance
[Fe/H], and metal abundance [$Z$/H] in the sample galaxies as a
function of the central velocity dispersion. The labels 
correspond to the row numbers in the Table
\ref{tab:GalProperties1}. The open and filled circles refer to
quiescent and active galaxies, respectively. The small and large
circles refer to galaxies with an age between 5 and 10\,Gyr and
$>$10\,Gyr, respectively. Galaxies with no known ages were assumed to
be old. The dashed lines corresponds to the relations find by
S08. The solid line in each panel represents the linear
regression (y = ax + b) through all the data points except for \#4,
\#12 and \#14. The Pearson correlation coefficient (r) and the results
of the linear fit are given. Typical error bars are shown.}
\label{fig:VelDips}
\end{figure}

The EW of our NIR spectral indices, the iron
abundance and metallicity are plotted versus the central velocity
dispersion in Fig.\,\ref{fig:VelDips}. The loci of the S08 data (see
their Fig. 13) are shown with dashed lines. Among the galaxies,
NGC\,5077 (\#12) and NGC\,6909 (\#14) possess very weak Na{\sc I},
Ca{\sc I} lines, and their relative errors are significant. The
correlations between the NIR indices and velocity dispersion for the
sample galaxies are plotted in Fig\,\ref{fig:VelDips}; the fit do not
consider NGC\,4281 (\#4), NGC\,5077 (\#12) and NGC\,6909 (\#14)
because of the reasons explained further on.  The correlation with CO
shows a similar slope as the one found by S08 but with a different
zero-point. The different definition of the index and especially the
position of the continuum bandpass can lead to a systematic variation
of the EW values. This effect seems to be more evident in the CO band,
whose equivalent width is measured with only extrapolated continuum
blue-wards of the feature.  The agreement with S08 is closer in the
Na{\sc I} versus $\sigma$ relation although our data show bigger
scatter. Without pretending to trace a general conclusion we found
relatively large NaI values compared to stars and something similar
was also found by S08. The iron abundance [Fe/H] and the total
abundance [Z/H] only show very weak correlations. Active galaxies are
clustered at the high end of the velocity dispersion distribution
which is understandable because their velocity dispersion measurement
may be affected by the black hole and the active nucleus. Note that
our highest velocity dispersion galaxy is a Type 2 Seyfert and the CO
band may suffer some dilution, as discussed in \citet{iva00}.

\begin{figure}
\includegraphics[width=9truecm]{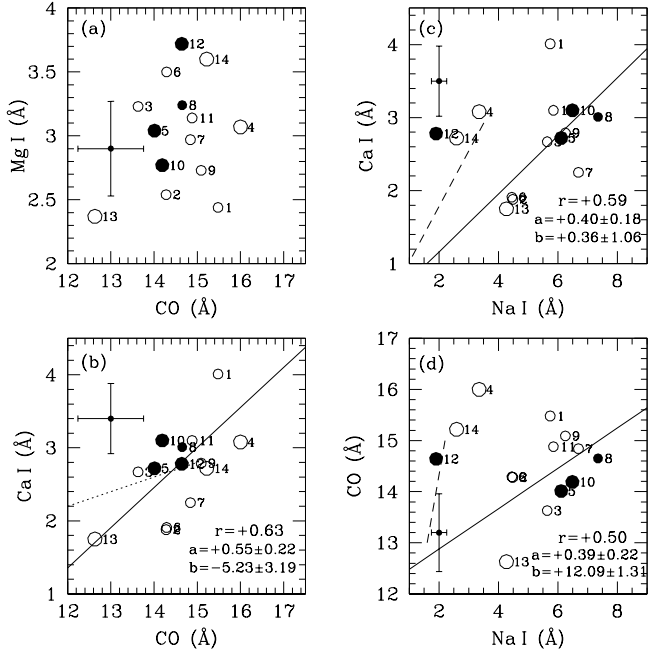}
\caption{The equivalent widths of the NIR indices plotted against each other.
The symbols are the same as in Fig.\,\ref{fig:VelDips}. The dashed
lines correspond to the relations found by S08. The dotted line
corresponds to the relation by \citet{dav08} for giant stars in the
solar neighborhoods. All these relations are plotted in their observed
range. The solid line represents the linear regression (y = ax
+ b) through all the data points except for \#4,
\#12 and \#14. The Pearson correlation coefficient (r) and the results
of the linear fit are given. Typical error bars are shown.}
\label{fig:IR_IR}
\end{figure}

The EW of our NIR spectral indices are plotted against each other in
Fig.\,\ref{fig:IR_IR} and compared to the relations for cluster stars
and solar neighborhood giants by S08 and \citet{dav08},
respectively. Three galaxies -- NGC\,4281 (\#4), NGC\,5077 (\#12) and
NGC\,6909 (\#14) -- are above the relation of S08 for the cluster
stars, which hints at different chemical enrichment history with
respect to the rest of the sample. Unfortunately the literature lacks
much data about these objects and although NGC\,6909 has the lowest
velocity dispersion in our sample, these galaxies do not stand out in
any respect, including in the optical Mg$_2$ versus $\sigma$ diagram.
Further investigation of these galaxies is necessary. The rest of our
objects populates a locus that follows similar trend as the galaxies
of S08. In the Ca{\sc I} versus CO plot we see a correlation with
significant scatter, and with a different slope than in S08.  Finally,
the Mg{\sc I} versus CO plot is dominated by scatter.  Galaxies with
traces of nuclear activity, evident in other wavelength ranges than
the NIR, do not seem to separate from the rest of the sample, and they
show no traces of emission lines. Therefore, the contribution of their
AGNs is negligible with respect to the rest of the galaxy.

\begin{figure}
\includegraphics[width=9truecm]{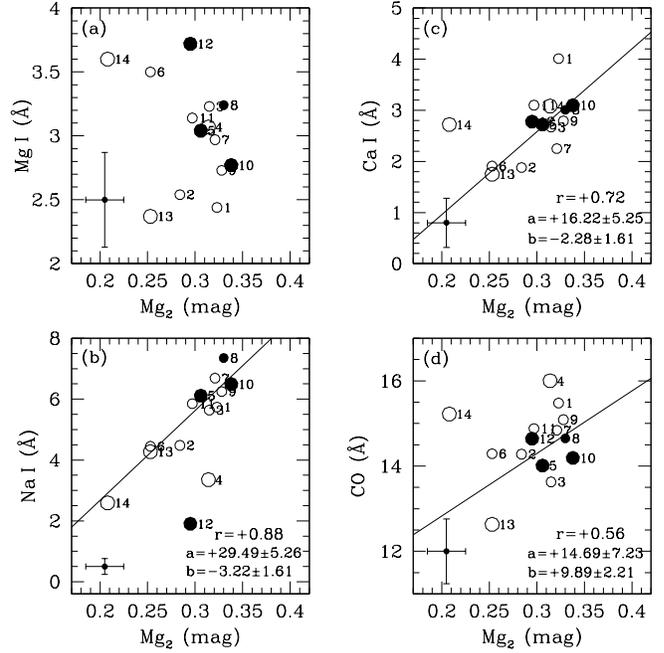}
\caption{The equivalent widths of the NIR indices plotted against the 
Mg$_2$ indices measured by \citet{ben93}. The symbols are the same as
in Fig.\,\ref{fig:VelDips}. The solid line represents the
linear regression (y = ax + b) through all the data points except for
\#4, \#12 and \#14. The Pearson correlation coefficient (r) and the results
of the linear fit are given. Typical error bars are shown.}
\label{fig:IR_FeH}
\end{figure}

Figure\,\ref{fig:IR_FeH} shows the EW of the NIR spectral indices
versus the Mg$_2$ measurements from \citet{ben93}. Na{\sc I}, and to a
lesser extent Ca{\sc I} and CO, do show trends with respect to Mg$_2$,
with NGC\,4281 (\#4), NGC\,5077 (\#12) and NGC\,6909 (\#14) standing
out. The correlation of Na{\sc I} with both $\sigma$ and Mg$_2$
suggests that the Na{\sc I} feature is dominated by the stellar
photosphere rather than by the interstellar medium.

\begin{figure}
\includegraphics[width=9truecm]{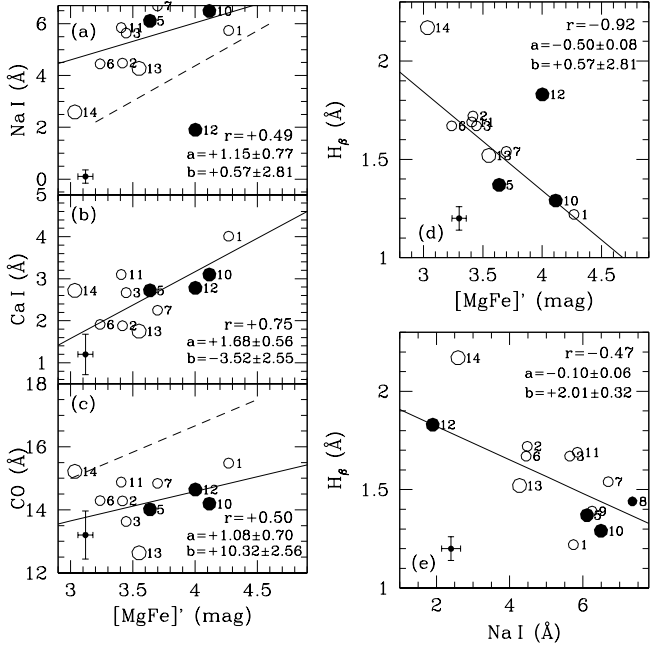}
\caption{(a-d) The equivalent widths of the NaI, CaI, CO and 
H$\beta$ indices plotted against the [MgFe]$''$ index. (e) The
equivalent widths of the H$\beta$ index as a function of the NaI
index. The symbols are the same as in Fig.\,\ref{fig:VelDips}. The
dashed lines correspond to the relations found by S08. 
The solid line represents the linear
regression (y = ax + b) through all the data points except for \#4,
\#12 and \#14. The Pearson correlation coefficient (r) and the results
of the linear fit are given. Typical error bars are shown.}
\label{fig:IR_MgFe}
\end{figure}

Given that [Fe/H] measurements are not available for the entire sample
and that Mg$_2$ is not representative of the total chemical abundance
of a galaxy because of the varying abundance ratios, we attempted to
create a combined Iron-and-$\alpha$-element index similar to that
defined by \citet{gon93} and more recently by \citet{tom03}:
\begin{equation}
{\rm [MgFe]'} = \sqrt{{\rm Mg}b\times (0.72 \times {\rm Fe}5270+0.28 \times {\rm Fe}5335)} 
\end{equation}
Such a combined indicator is expected to decrease the effect from the
varying $\alpha$/Fe ratio. Since not all the components of
this indicator are available we directly substitute the iron and
magnesium indices by defining a new indicator:
\begin{equation}\label{Eqn:FeMg2}
{\rm [MgFe]''} = \sqrt{{\rm Mg}_2\times {\rm Fe}5335}
\end{equation}
The results are shown in Fig.\,\ref{fig:IR_MgFe}a-c.  The NaI index of
NGC\,5077 (\#12) and the CO index of NGC\,6909 (\#14) deviate from the
main loci of the other galaxies.

To address this issue we plotted the optical indices of these galaxies
versus the combined [MgFe]$'$ index (Fig.\,\ref{fig:IR_MgFe}d). This
is an analog of the typical plot \citep[i.e.][]{wor94} that allows to
disentangle the age-metallicity degeneracy: the inverse \Hb index is
roughly proportional to age while the [MgFe]$'$ is dominated by metal
abundance. The two galaxies exhibit strong \Hb which suggests that
they are dominated by populations of 3\,Gyr or younger (see for
example Fig. 1 in S08). NGC\,5077 and NGC\,6909 are also well
separated from the rest of the galaxies on the \Hb versus Na{\sc I}
plot (Fig.\,\ref{fig:IR_MgFe}e) which leads us to the conclusion that
the NIR indices can be used to create a similar diagnostic plot, but
measuring the Br$\gamma$ feature requires better quality data than the
ones described here.\\
We investigated the behavior of the NIR indices in relation to the H 
and K-band magnitude and the H-K color but no clear correlations were found.

\subsection{Combined NIR metal index}

\begin{figure}
\includegraphics[width=9truecm]{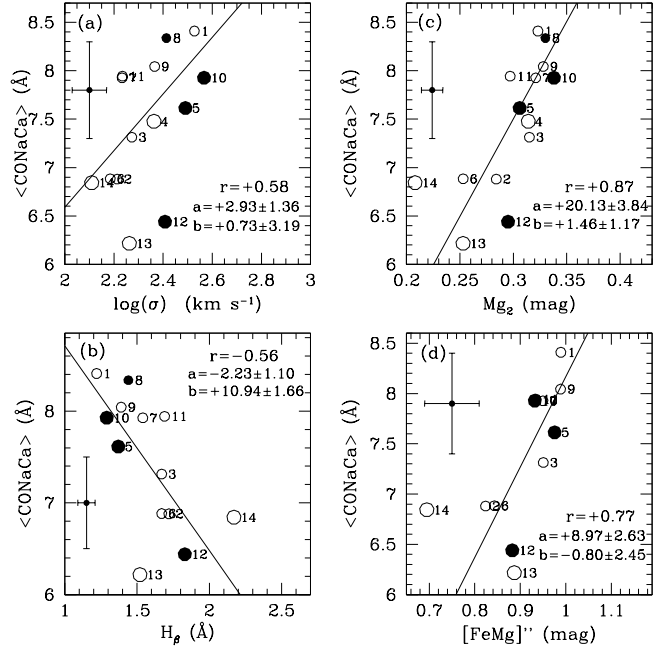}
\caption{The equivalent widths of the newly defined $<$CONaCa$>$ as a function of
the central velocity dispersion (a), Mg$_2$ (b), H$\beta$ (c), and
[FeMg]$''$ (d) indices. The symbols are the same as in
Fig.\,\ref{fig:VelDips}. The solid line represents the linear
regression (y = ax + b) through all the data points except for \#4,
\#12 and \#14. The Pearson correlation coefficient (r) and the results
of the linear fit are given. Typical error bars are shown.}
\label{fig:CONaCa}
\end{figure}

Ground based NIR spectroscopy is much more time consuming than the
corresponding optical observations -- higher and variable background
and detectors with worse cosmetics often require one to sacrifice
either resolution or signal-to-noise to obtain the data in a
reasonable amount of time. To alleviate this problem, we defined a
combined spectral index of all the major $K$-band metal features:
\begin{equation}
\langle {\rm CONaCa} \rangle = ( {\rm CO} + \ion{Na}{i} + \ion{Ca}{i} ) / 3.0 
\end{equation}
Various weighting schemes were tried to minimize the scatter of the
basic relations (Fig.\,\ref{fig:CONaCa}). However, the simple average
yielded the tightest relations. The carbon and oxygen are $\alpha$
elements, while the sodium and calcium originate in both high and low
mass stars. The average value of the CO index for the sample galaxies
is $\sim$15\,\AA, the Na{\sc I} is $\sim$4.3\,\AA\, and the Ca{\sc I}
is $\sim$2.7\,\AA\, which means that at least 2/3 of the new index is
dominated by metals produced mostly in high mass stars, i.e. early in
the history of the elliptical galaxies. The other implication is that
a relatively limited amount of recent star formation could affect the
new index more than it would an iron peak dominated index. The lack of
iron peak features in the NIR is well known as it was pointed out in
\citet{iva01}.

The new index improves the correlations. For example, the Pearson correlation 
coefficient is $\sim$10\% higher for the $\langle {\rm CONaCa}
\rangle$ vs. Mg2 and even $\sim$20\% higher for the $\langle {\rm
CONaCa} \rangle$ vs. [MgFe]'', with respect to the respective
correlations where only one IR index is used. We derived these
relations using only the galaxies with known parameters,
i.e. excluding the poorly studied NGC\,4281(\#4) and, NGC\,5077(\#12)
and NGC\,6909(\#14) which systematically show peculiarities with
respect to the bulk of our sample.

\subsection{Average NIR spectrum of the sample galaxies}
\begin{figure}
\includegraphics[width=9truecm]{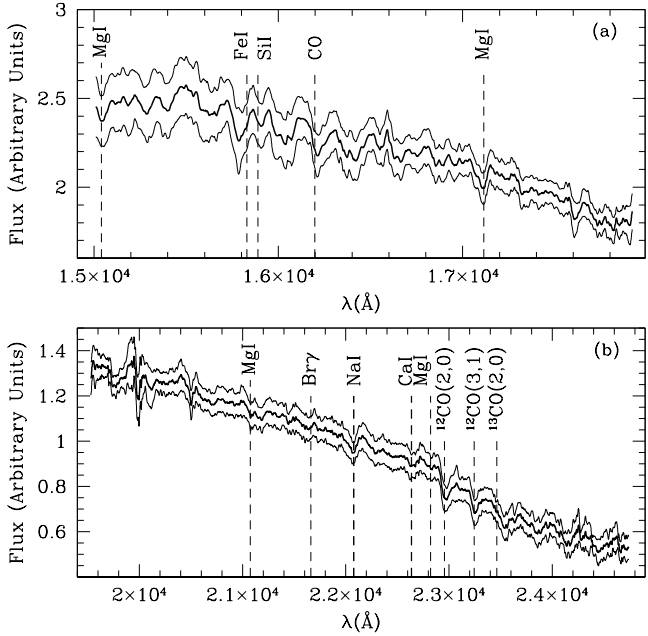}
\caption{Average $H$- (a) and $K-$band (b) spectrum of sample galaxies
normalized to unity at 22000\ \AA\, (thick line). The thin lines
correspond to the $\pm1\sigma$ confidence levels. Some of the more
prominent spectral features are marked.}
\label{fig:AverSpec}
\end{figure}

Studies of composite stellar systems (i.e. galaxies hosting an AGN) 
often need to subtract the contribution of the underlying galaxy.
This prompted us to create an average spectrum
of the galaxies in our sample. We used a homogenized subset of eight
galaxies, excluding the objects with young populations. We also
excluded those with low signal-to-noise ratio. The average age of the
galaxies used to create the composite spectrum is 9$\pm$2\,Gyr (the
median is 9.5\,Gyr) and the average [Fe/H] is 0.17$\pm$0.13 (the
median is 0.16).  The composite spectrum is shown in
Fig.\,\ref{fig:AverSpec} and the values of the spectrum are listed in
Table\,\ref{tab:AverSpec}, together with the r.m.s. values per
 Angstrom.  We included the $H$-band section because the
artifacts caused by the spectral type mismatch of the standards are
minimized by the averaging of galaxies observed at different
redshifts.

\begin{table}
\caption{Average spectrum of the sample galaxies
observed at NTT in arbitrary units and normalized to unity at
22000\,\AA. The regions with zero r.m.s. at the beginning and the at
the and are covered only by one or two spectra. The full table is
given only in the electronic version of the journal.}
\label{tab:AverSpec}
\begin{center}
\begin{tabular}{ccc}
\hline
\hline
$\lambda$ (\AA) & F$_\lambda$ & r.m.s. \\
\hline
14944 & 2.217 & 0.000 \\
14945 & 2.384 & 0.000 \\
14946 & 2.384 & 0.000 \\
14947 & 2.386 & 0.000 \\
14948 & 2.130 & 0.366 \\
14949 & 2.572 & 0.254 \\
14950 & 2.575 & 0.251 \\
14951 & 2.578 & 0.248 \\
14952 & 2.583 & 0.245 \\
14953 & 2.588 & 0.243 \\
14954 & 2.595 & 0.241 \\
14955 & 2.602 & 0.240 \\
\hline
\end{tabular}
\end{center}
\end{table}

\subsection{The Mg{\sc I} feature at 1.51\,$\mu$m}

\begin{figure}
\includegraphics[width=9truecm]{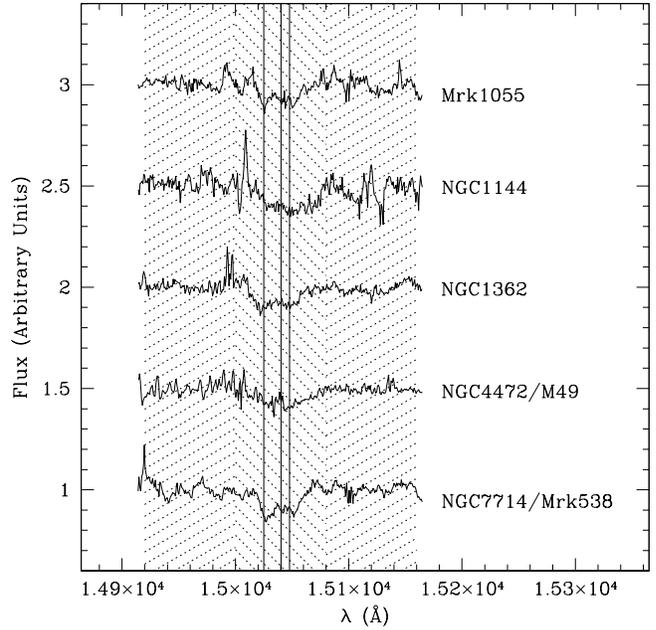}
\caption{Spectra of the galaxies observed with ISAAC in the region of 
the Mg{\sc I} (1.51 $\mu$m). The central wavelengths of the individual
Mg{\sc I} lines are shown with vertical solid lines.  The passbands
for the feature and the adjacent continua are shaded. The spectra were
normalized to unity and shifted vertically by 0.5 for display
purposes.  }
\label{fig:MgI_H_band}
\end{figure}

\citet{iva04} pointed out the possibility of using the
Mg{\sc I} feature at 1.51\,$\mu$m as an $\alpha$-element abundance
indicator in the NIR. With exploratory purposes we obtained spectra of
a small and diverse set of galaxies (Table\,\ref{tab:log1} and
Fig.\,\ref{fig:MgI_H_band}). To carry out quantitative analysis we
defined an index (Table\,\ref{tab:bandDef}) and for the five galaxies
we measured equivalent widths of 3.3, 5.0, 3.8, 3.9 and 4.5\AA,
respectively for Mrk\,1055, NGC\,1144, NGC\,1362, NGC\,4472 and
NGC\,7714. The typical uncertainty is $\sim0.3$\AA. No corrections for
velocity dispersion were applied. Given the diverse nature of the
objects, we refrain from drawing any conclusions but we note that the
relation between the optical and the NIR Mg features is not
straightforward because the only two Mg$_2$ values that we have from
the literature are inversely proportional to our measurements for
NGC\,1362 and NGC\,4472. Nevertheless, these observations prove that
it is feasible to measure the NIR Mg{\sc I} feature at 1.51\,$\mu$m
and give a basis for future NIR synthetic spectral modeling.

%____________________________________________________________________
\section{Summary\label{sec:Results}}
\begin{enumerate}
\item We report on a new data set of NIR spectra of ellipticals/spheroids. 
The stronger features were measured and show mild correlations with
the optical metal features. NIR versus NIR features show a lot of
scatter but the diagram of CO versus Na{\sc I} seems to be a good
diagnostic for detecting different chemical enrichment histories.
\item  The NIR metal absorptions, in combination with a Hydrogen 
absorption feature may be able to break the age-metallicity degeneracy,
but further investigation and better data are needed to verify this
possibility.
\item A new combined NIR index is defined that allows one to interpret lower
signal-to-noise data. It is dominated, at about 2/3 level, by
$\alpha$-elements. We also present an average spectrum of
ellipticals/spheroids, built from a homogenized subset of our sample
that can be used as template of the integrated early-type galaxy
spectra.
\item The strength of Mg{\sc I} feature at 1.5\,$\mu$m was measured for the
first time in a heterogeneous sample of galaxies, providing a new
constraint for future NIR spectral synthesis models.
\end{enumerate}
\begin{acknowledgements}
We thank Reynier Peletier for useful discussion and suggestions. We
acknowledge grant CPDA068415/06 by Padua University, which provided
support for this research. MC thanks the European Southern Observatory
for support via ESO Studentship program. LM is supported by grant
(CPDR061795/06) by Padova University. LM acknowledge the Pontificia
Universidad Catolica de Chile and the European Southern Observatory
for hospitality while this paper was in progress. We are grateful to
the La Silla and Paranal staff astronomers for their support.
\end{acknowledgements}

\bibliographystyle{aa}
\bibliography{IRspectra.bib}

\begin{thebibliography}{81}
\expandafter\ifx\csname natexlab\endcsname\relax\def\natexlab#1{#1}\fi

\bibitem[{{Ali} {et~al.}(1995){Ali}, {Carr}, {Depoy}, {Frogel}, \&
  {Sellgren}}]{ali95}
{Ali}, B., {Carr}, J.~S., {Depoy}, D.~L., {Frogel}, J.~A., \& {Sellgren}, K.
  1995, \aj, 110, 2415

\bibitem[{{Alonso-Herrero} {et~al.}(2000){Alonso-Herrero}, {Rieke}, {Rieke}, \&
  {Shields}}]{alo00}
{Alonso-Herrero}, A., {Rieke}, M.~J., {Rieke}, G.~H., \& {Shields}, J.~C. 2000,
  \apj, 530, 688

\bibitem[{{Annibali} {et~al.}(2007){Annibali}, {Bressan}, {Rampazzo},
  {Zeilinger}, \& {Danese}}]{ann07}
{Annibali}, F., {Bressan}, A., {Rampazzo}, R., {Zeilinger}, W.~W., \& {Danese},
  L. 2007, \aap, 463, 455

\bibitem[{{Bender} {et~al.}(1993){Bender}, {Burstein}, \& {Faber}}]{ben93}
{Bender}, R., {Burstein}, D., \& {Faber}, S.~M. 1993, \apj, 411, 153

\bibitem[{{Bernardi} {et~al.}(2002){Bernardi}, {Alonso}, {da Costa}, {Willmer},
  {Wegner}, {Pellegrini}, {Rit{\'e}}, \& {Maia}}]{ber02}
{Bernardi}, M., {Alonso}, M.~V., {da Costa}, L.~N., {et~al.} 2002, \aj, 123,
  2990

\bibitem[{{Bernardi} {et~al.}(1998){Bernardi}, {Renzini}, {da Costa}, {Wegner},
  {Alonso}, {Pellegrini}, {Rit{\'e}}, \& {Willmer}}]{ber98}
{Bernardi}, M., {Renzini}, A., {da Costa}, L.~N., {et~al.} 1998, \apjl, 508,
  L143

\bibitem[{{Binggeli} {et~al.}(1985){Binggeli}, {Sandage}, \& {Tammann}}]{bin85}
{Binggeli}, B., {Sandage}, A., \& {Tammann}, G.~A. 1985, \aj, 90, 1681

\bibitem[{{Boisson} {et~al.}(2002){Boisson}, {Coup{\'e}}, {Cuby}, {Joly}, \&
  {Ward}}]{boi02}
{Boisson}, C., {Coup{\'e}}, S., {Cuby}, J.~G., {Joly}, M., \& {Ward}, M.~J.
  2002, \aap, 396, 489

\bibitem[{{Burstein} {et~al.}(1984){Burstein}, {Faber}, {Gaskell}, \&
  {Krumm}}]{bur84}
{Burstein}, D., {Faber}, S.~M., {Gaskell}, C.~M., \& {Krumm}, N. 1984, \apj,
  287, 586

\bibitem[{{Burston} {et~al.}(2001){Burston}, {Ward}, \& {Davies}}]{bur01}
{Burston}, A.~J., {Ward}, M.~J., \& {Davies}, R.~I. 2001, \mnras, 326, 403

\bibitem[{{Cenarro} {et~al.}(2003){Cenarro}, {Gorgas}, {Vazdekis}, {Cardiel},
  \& {Peletier}}]{cen03}
{Cenarro}, A.~J., {Gorgas}, J., {Vazdekis}, A., {Cardiel}, N., \& {Peletier},
  R.~F. 2003, \mnras, 339, L12

\bibitem[{{Coziol} {et~al.}(2001){Coziol}, {Doyon}, \& {Demers}}]{coz01}
{Coziol}, R., {Doyon}, R., \& {Demers}, S. 2001, \mnras, 325, 1081

\bibitem[{{Cutri} {et~al.}(2003){Cutri}, {Skrutskie}, {van Dyk}, {Beichman},
  {Carpenter}, {Chester}, {Cambresy}, {Evans}, {Fowler}, {Gizis}, {Howard},
  {Huchra}, {Jarrett}, {Kopan}, {Kirkpatrick}, {Light}, {Marsh}, {McCallon},
  {Schneider}, {Stiening}, {Sykes}, {Weinberg}, {Wheaton}, {Wheelock}, \&
  {Zacarias}}]{cut03}
{Cutri}, R.~M., {Skrutskie}, M.~F., {van Dyk}, S., {et~al.} 2003, {2MASS All
  Sky Catalog of point sources.} (The IRSA 2MASS All-Sky Point Source Catalog,
  NASA/IPAC Infrared Science
  Archive.~http://irsa.ipac.caltech.edu/applications/Gator/)

\bibitem[{{da Costa} {et~al.}(1998){da Costa}, {Willmer}, {Pellegrini},
  {Chaves}, {Rit{\'e}}, {Maia}, {Geller}, {Latham}, {Kurtz}, {Huchra},
  {Ramella}, {Fairall}, {Smith}, \& {L{\'{\i}}pari}}]{dac98}
{da Costa}, L.~N., {Willmer}, C.~N.~A., {Pellegrini}, P.~S., {et~al.} 1998,
  \aj, 116, 1

\bibitem[{{Dalle Ore} {et~al.}(1991){Dalle Ore}, {Faber}, {Jesus}, {Stoughton},
  \& {Burstein}}]{dal91}
{Dalle Ore}, C., {Faber}, S.~M., {Jesus}, J., {Stoughton}, R., \& {Burstein},
  D. 1991, \apj, 366, 38

\bibitem[{{Dallier} {et~al.}(1996){Dallier}, {Boisson}, \& {Joly}}]{dal96}
{Dallier}, R., {Boisson}, C., \& {Joly}, M. 1996, \aaps, 116, 239

\bibitem[{{Das} \& {Jog}(1999)}]{das99}
{Das}, M. \& {Jog}, C.~J. 1999, \apj, 527, 600

\bibitem[{{Davidge} {et~al.}(2008){Davidge}, {Beck}, \& {McGregor}}]{dav08}
{Davidge}, T.~J., {Beck}, T.~L., \& {McGregor}, P.~J. 2008, \apj, 677, 238

\bibitem[{{de Vaucouleurs} {et~al.}(1991){de Vaucouleurs}, {de Vaucouleurs},
  {Corwin}, {Buta}, {Paturel}, \& {Fouque}}]{dev91}
{de Vaucouleurs}, G., {de Vaucouleurs}, A., {Corwin}, Jr., H.~G., {et~al.}
  1991, {Third Reference Catalogue of Bright Galaxies} (Springer-Verlag,
  Berlin)

\bibitem[{{Denicol{\'o}} {et~al.}(2005){Denicol{\'o}}, {Terlevich},
  {Terlevich}, {Forbes}, {Terlevich}, \& {Carrasco}}]{den05}
{Denicol{\'o}}, G., {Terlevich}, R., {Terlevich}, E., {et~al.} 2005, \mnras,
  356, 1440

\bibitem[{{Devillard}(1999)}]{Devillard1999}
{Devillard}, N. 1999, in Astronomical Society of the Pacific Conference Series,
  Vol. 172, Astronomical Data Analysis Software and Systems VIII, ed. D.~M.
  {Mehringer}, R.~L. {Plante}, \& D.~A. {Roberts}, 333

\bibitem[{{Engelbracht}(1997)}]{eng97}
{Engelbracht}, C.~W. 1997, PhD thesis, Univ. of Arizona

\bibitem[{{Faber}(1973)}]{fab73}
{Faber}, S.~M. 1973, \apj, 179, 731

\bibitem[{{Faber} {et~al.}(1985){Faber}, {Friel}, {Burstein}, \&
  {Gaskell}}]{fab85}
{Faber}, S.~M., {Friel}, E.~D., {Burstein}, D., \& {Gaskell}, C.~M. 1985,
  \apjs, 57, 711

\bibitem[{{Faber} {et~al.}(1999){Faber}, {Trager}, {Gonzalez}, \&
  {Worthey}}]{fab99}
{Faber}, S.~M., {Trager}, S.~C., {Gonzalez}, J.~J., \& {Worthey}, G. 1999,
  \apss, 267, 273

\bibitem[{{Falco} {et~al.}(1999){Falco}, {Kurtz}, {Geller}, {Huchra}, {Peters},
  {Berlind}, {Mink}, {Tokarz}, \& {Elwell}}]{fal99}
{Falco}, E.~E., {Kurtz}, M.~J., {Geller}, M.~J., {et~al.} 1999, \pasp, 111, 438

\bibitem[{{F{\"o}rster Schreiber}(2000)}]{for00}
{F{\"o}rster Schreiber}, N.~M. 2000, \aj, 120, 2089

\bibitem[{{Frogel} {et~al.}(2001){Frogel}, {Stephens}, {Ram{\'{\i}}rez}, \&
  {DePoy}}]{fro01}
{Frogel}, J.~A., {Stephens}, A., {Ram{\'{\i}}rez}, S., \& {DePoy}, D.~L. 2001,
  \aj, 122, 1896

\bibitem[{{Gardner} {et~al.}(2006){Gardner}, {Mather}, {Clampin}, {Doyon},
  {Greenhouse}, {Hammel}, {Hutchings}, {Jakobsen}, {Lilly}, {Long}, {Lunine},
  {McCaughrean}, {Mountain}, {Nella}, {Rieke}, {Rieke}, {Rix}, {Smith},
  {Sonneborn}, {Stiavelli}, {Stockman}, {Windhorst}, \& {Wright}}]{gar06}
{Gardner}, J.~P., {Mather}, J.~C., {Clampin}, M., {et~al.} 2006, Space Science
  Reviews, 123, 485

\bibitem[{{Goldader} {et~al.}(1995){Goldader}, {Joseph}, {Doyon}, \&
  {Sanders}}]{gol95}
{Goldader}, J.~D., {Joseph}, R.~D., {Doyon}, R., \& {Sanders}, D.~B. 1995,
  \apj, 444, 97

\bibitem[{{Goldader} {et~al.}(1997){Goldader}, {Joseph}, {Doyon}, \&
  {Sanders}}]{gol97}
{Goldader}, J.~D., {Joseph}, R.~D., {Doyon}, R., \& {Sanders}, D.~B. 1997,
  \apjs, 108, 449

\bibitem[{{Gonz{\'a}lez}(1993)}]{gon93}
{Gonz{\'a}lez}, J.~J. 1993, PhD thesis, Univ. of Califonia, Santa Cruz

\bibitem[{{Hanson} {et~al.}(1996){Hanson}, {Conti}, \& {Rieke}}]{han96}
{Hanson}, M.~M., {Conti}, P.~S., \& {Rieke}, M.~J. 1996, \apjs, 107, 281

\bibitem[{{Heckman}(1980)}]{hec80}
{Heckman}, T.~M. 1980, \aap, 87, 152

\bibitem[{{Ho} {et~al.}(1997){Ho}, {Filippenko}, \& {Sargent}}]{ho97}
{Ho}, L.~C., {Filippenko}, A.~V., \& {Sargent}, W.~L.~W. 1997, \apjs, 112, 315

\bibitem[{{Howell}(2005)}]{how05}
{Howell}, J.~H. 2005, \aj, 130, 2065

\bibitem[{{Ivanov}(2001)}]{iva01}
{Ivanov}, V.~D. 2001, PhD thesis, Univ. of Arizona

\bibitem[{{Ivanov} {et~al.}(2000){Ivanov}, {Rieke}, {Groppi}, {Alonso-Herrero},
  {Rieke}, \& {Engelbracht}}]{iva00}
{Ivanov}, V.~D., {Rieke}, G.~H., {Groppi}, C.~E., {et~al.} 2000, \apj, 545, 190

\bibitem[{{Ivanov} {et~al.}(2004){Ivanov}, {Rieke}, {Engelbracht},
  {Alonso-Herrero}, {Rieke}, \& {Luhman}}]{iva04}
{Ivanov}, V.~D., {Rieke}, M.~J., {Engelbracht}, C.~W., {et~al.} 2004, \apjs,
  151, 387

\bibitem[{{Keel}(1996)}]{kee96}
{Keel}, W.~C. 1996, \apjs, 106, 27

\bibitem[{{Kleinmann} \& {Hall}(1986)}]{kle86}
{Kleinmann}, S.~G. \& {Hall}, D.~N.~B. 1986, \apjs, 62, 501

\bibitem[{{Kobayashi} \& {Arimoto}(1999)}]{kob99}
{Kobayashi}, C. \& {Arimoto}, N. 1999, \apj, 527, 573

\bibitem[{{Lancon} \& {Rocca-Volmerange}(1992)}]{lan92}
{Lancon}, A. \& {Rocca-Volmerange}, B. 1992, \aaps, 96, 593

\bibitem[{{Larkin} {et~al.}(1998){Larkin}, {Armus}, {Knop}, {Soifer}, \&
  {Matthews}}]{lar98}
{Larkin}, J.~E., {Armus}, L., {Knop}, R.~A., {Soifer}, B.~T., \& {Matthews}, K.
  1998, \apjs, 114, 59

\bibitem[{{Lauberts} \& {Valentijn}(1989)}]{lau89}
{Lauberts}, A. \& {Valentijn}, E.~A. 1989, The Messenger, 56, 31

\bibitem[{{Lu} {et~al.}(1993){Lu}, {Hoffman}, {Groff}, {Roos}, \&
  {Lamphier}}]{lu93}
{Lu}, N.~Y., {Hoffman}, G.~L., {Groff}, T., {Roos}, T., \& {Lamphier}, C. 1993,
  \apjs, 88, 383

\bibitem[{{Maiolino} {et~al.}(1996){Maiolino}, {Rieke}, \& {Rieke}}]{mai96}
{Maiolino}, R., {Rieke}, G.~H., \& {Rieke}, M.~J. 1996, \aj, 111, 537

\bibitem[{{Mannucci} {et~al.}(2001){Mannucci}, {Basile}, {Poggianti},
  {Cimatti}, {Daddi}, {Pozzetti}, \& {Vanzi}}]{man01}
{Mannucci}, F., {Basile}, F., {Poggianti}, B.~M., {et~al.} 2001, \mnras, 326,
  745

\bibitem[{{Mehlert} {et~al.}(2003){Mehlert}, {Thomas}, {Saglia}, {Bender}, \&
  {Wegner}}]{meh03}
{Mehlert}, D., {Thomas}, D., {Saglia}, R.~P., {Bender}, R., \& {Wegner}, G.
  2003, \aap, 407, 423

\bibitem[{{Meyer} {et~al.}(1998){Meyer}, {Edwards}, {Hinkle}, \&
  {Strom}}]{mey98}
{Meyer}, M.~R., {Edwards}, S., {Hinkle}, K.~H., \& {Strom}, S.~E. 1998, \apj,
  508, 397

\bibitem[{{Moorwood} {et~al.}(1998{\natexlab{a}}){Moorwood}, {Cuby},
  {Biereichel}, {Brynnel}, {Delabre}, {Devillard}, {van Dijsseldonk}, {Finger},
  {Gemperlein}, {Gilmozzi}, {Herlin}, {Huster}, {Knudstrup}, {Lidman}, {Lizon},
  {Mehrgan}, {Meyer}, {Nicolini}, {Petr}, {Spyromilio}, \&
  {Stegmeier}}]{moo98a}
{Moorwood}, A., {Cuby}, J.-G., {Biereichel}, P., {et~al.} 1998{\natexlab{a}},
  The Messenger, 94, 7

\bibitem[{{Moorwood} {et~al.}(1998{\natexlab{b}}){Moorwood}, {Cuby}, \&
  {Lidman}}]{moo98b}
{Moorwood}, A., {Cuby}, J.-G., \& {Lidman}, C. 1998{\natexlab{b}}, The
  Messenger, 91, 9

\bibitem[{{Morelli} {et~al.}(2008){Morelli}, {Pompei}, {Pizzella},
  {M{\'e}ndez-Abreu}, {Corsini}, {Coccato}, {Saglia}, {Sarzi}, \&
  {Bertola}}]{mor08}
{Morelli}, L., {Pompei}, E., {Pizzella}, A., {et~al.} 2008, {MNRAS, in press
  (arXiv:0806.2988)}

\bibitem[{{Murphy} {et~al.}(2001){Murphy}, {Soifer}, {Matthews}, {Armus}, \&
  {Kiger}}]{mur01}
{Murphy}, Jr., T.~W., {Soifer}, B.~T., {Matthews}, K., {Armus}, L., \& {Kiger},
  J.~R. 2001, \aj, 121, 97

\bibitem[{{Murphy} {et~al.}(1999){Murphy}, {Soifer}, {Matthews}, {Kiger}, \&
  {Armus}}]{mur99}
{Murphy}, Jr., T.~W., {Soifer}, B.~T., {Matthews}, K., {Kiger}, J.~R., \&
  {Armus}, L. 1999, \apjl, 525, L85

\bibitem[{{O'Connell}(1986)}]{oco86}
{O'Connell}, R.~W. 1986, \pasp, 98, 163

\bibitem[{{Ogando} {et~al.}(2008){Ogando}, {Maia}, {Pellegrini}, \& {da
  Costa}}]{oga08}
{Ogando}, R.~L.~C., {Maia}, M.~A.~G., {Pellegrini}, P.~S., \& {da Costa}, L.~N.
  2008, \aj, 135, 2424

\bibitem[{{Oliva} {et~al.}(1995){Oliva}, {Origlia}, {Kotilainen}, \&
  {Moorwood}}]{oli95}
{Oliva}, E., {Origlia}, L., {Kotilainen}, J.~K., \& {Moorwood}, A.~F.~M. 1995,
  \aap, 301, 55

\bibitem[{{Origlia} {et~al.}(1993){Origlia}, {Moorwood}, \& {Oliva}}]{ori93}
{Origlia}, L., {Moorwood}, A.~F.~M., \& {Oliva}, E. 1993, \aap, 280, 536

\bibitem[{{Pickles}(1998)}]{pic98}
{Pickles}, A.~J. 1998, \pasp, 110, 863

\bibitem[{{Pych}(2004)}]{pyc04}
{Pych}, W. 2004, \pasp, 116, 148

\bibitem[{{Rampazzo} {et~al.}(2005){Rampazzo}, {Annibali}, {Bressan},
  {Longhetti}, {Padoan}, \& {Zeilinger}}]{ram05}
{Rampazzo}, R., {Annibali}, F., {Bressan}, A., {et~al.} 2005, \aap, 433, 497

\bibitem[{{Reunanen} {et~al.}(2002){Reunanen}, {Kotilainen}, \&
  {Prieto}}]{reu02}
{Reunanen}, J., {Kotilainen}, J.~K., \& {Prieto}, M.~A. 2002, \mnras, 331, 154

\bibitem[{{Reunanen} {et~al.}(2007){Reunanen}, {Tacconi-Garman}, \&
  {Ivanov}}]{reu07}
{Reunanen}, J., {Tacconi-Garman}, L.~E., \& {Ivanov}, V.~D. 2007, \mnras, 382,
  951

\bibitem[{{Rinn} {et~al.}(2005){Rinn}, {Sambruna}, \& {Gliozzi}}]{rin05}
{Rinn}, A.~S., {Sambruna}, R.~M., \& {Gliozzi}, M. 2005, \apj, 621, 167

\bibitem[{{S{\'a}nchez-Bl{\'a}zquez} {et~al.}(2007){S{\'a}nchez-Bl{\'a}zquez},
  {Forbes}, {Strader}, {Brodie}, \& {Proctor}}]{san07}
{S{\'a}nchez-Bl{\'a}zquez}, P., {Forbes}, D.~A., {Strader}, J., {Brodie}, J.,
  \& {Proctor}, R. 2007, \mnras, 377, 759

\bibitem[{{Silva} {et~al.}(2008){Silva}, {Kuntschner}, \& {Lyubenova}}]{sil08}
{Silva}, D.~R., {Kuntschner}, H., \& {Lyubenova}, M. 2008, \apj, 674, 194

\bibitem[{{Smith} {et~al.}(2000){Smith}, {Lucey}, {Hudson}, {Schlegel}, \&
  {Davies}}]{smi00}
{Smith}, R.~J., {Lucey}, J.~R., {Hudson}, M.~J., {Schlegel}, D.~J., \&
  {Davies}, R.~L. 2000, \mnras, 313, 469

\bibitem[{{Sosa-Brito} {et~al.}(2001){Sosa-Brito}, {Tacconi-Garman}, {Lehnert},
  \& {Gallimore}}]{sos01}
{Sosa-Brito}, R.~M., {Tacconi-Garman}, L.~E., {Lehnert}, M.~D., \& {Gallimore},
  J.~F. 2001, \apjs, 136, 61

\bibitem[{{Stephens} \& {Frogel}(2004)}]{ste04}
{Stephens}, A.~W. \& {Frogel}, J.~A. 2004, \aj, 127, 925

\bibitem[{{Terlevich} {et~al.}(1981){Terlevich}, {Davies}, {Faber}, \&
  {Burstein}}]{ter81}
{Terlevich}, R., {Davies}, R.~L., {Faber}, S.~M., \& {Burstein}, D. 1981,
  \mnras, 196, 381

\bibitem[{{Thomas} {et~al.}(2003){Thomas}, {Maraston}, \& {Bender}}]{tom03}
{Thomas}, D., {Maraston}, C., \& {Bender}, R. 2003, \mnras, 339, 897

\bibitem[{{Trager} {et~al.}(2000){Trager}, {Faber}, {Worthey}, \&
  {Gonz{\'a}lez}}]{tra00}
{Trager}, S.~C., {Faber}, S.~M., {Worthey}, G., \& {Gonz{\'a}lez}, J.~J. 2000,
  \aj, 120, 165

\bibitem[{{Trager} {et~al.}(1998){Trager}, {Worthey}, {Faber}, {Burstein}, \&
  {Gonzalez}}]{tra98}
{Trager}, S.~C., {Worthey}, G., {Faber}, S.~M., {Burstein}, D., \& {Gonzalez},
  J.~J. 1998, \apjs, 116, 1

\bibitem[{{Trager} {et~al.}(2005){Trager}, {Worthey}, {Faber}, \&
  {Dressler}}]{tra05}
{Trager}, S.~C., {Worthey}, G., {Faber}, S.~M., \& {Dressler}, A. 2005, \mnras,
  362, 2

\bibitem[{{van Dokkum}(2001)}]{vand01}
{van Dokkum}, P.~G. 2001, \pasp, 113, 1420

\bibitem[{{Vanzi} {et~al.}(1998){Vanzi}, {Alonso-Herrero}, \& {Rieke}}]{van98}
{Vanzi}, L., {Alonso-Herrero}, A., \& {Rieke}, G.~H. 1998, \apj, 504, 93

\bibitem[{{Vanzi} \& {Rieke}(1997)}]{van97}
{Vanzi}, L. \& {Rieke}, G.~H. 1997, \apj, 479, 694

\bibitem[{{Wallace} \& {Hinkle}(1997)}]{wal97}
{Wallace}, L. \& {Hinkle}, K. 1997, \apjs, 111, 445

\bibitem[{{Wegner} {et~al.}(2003){Wegner}, {Bernardi}, {Willmer}, {da Costa},
  {Alonso}, {Pellegrini}, {Maia}, {Chaves}, \& {Rit{\'e}}}]{weg03}
{Wegner}, G., {Bernardi}, M., {Willmer}, C.~N.~A., {et~al.} 2003, \aj, 126,
  2268

\bibitem[{{Worthey}(1994)}]{wor94}
{Worthey}, G. 1994, \apjs, 95, 107

\end{thebibliography}

\end{document}